\PassOptionsToPackage{colorlinks,citecolor=blue,urlcolor=blue}{hyperref}
\documentclass[11pt]{article}

\usepackage[margin=1in]{geometry}
\usepackage{lipsum}
\usepackage{amsthm,amsmath,amsfonts,amssymb}
\usepackage[authoryear]{natbib}
\usepackage{xcolor}
\usepackage{float}
\usepackage{booktabs}
\usepackage{tikz}
\usepackage{orcidlink}
\usetikzlibrary{bayesnet}
\usetikzlibrary{positioning}
\usepackage[colorlinks,citecolor=blue,urlcolor=blue]{hyperref}
\usepackage{graphicx}
\usepackage{url}

\usepackage{float}
\usepackage{algorithm}
\usepackage{algpseudocode}
\usepackage{empheq}
\usepackage{empheq}
\usepackage{booktabs}   
\usepackage{multirow}   
\usepackage{etoolbox}
\usepackage{bm}

\def\oper#1{\csdef{#1}{\operatorname{#1}}}
\forcsvlist\oper{GL,SL,deg,Hom,End,Vect,Mod,Rep,Ker,Im,Id}

\def\defbb#1{\csdef{b#1}{\mathbb{#1}}}
\forcsvlist\defbb{A,B,C,D,E,F,G,H,I,J,K,L,M,N,O,P,Q,R,S,T,U,V,W,X,Y,Z}

\def\defcal#1{\csdef{c#1}{\mathcal{#1}}}
\forcsvlist\defcal{A,B,C,D,E,F,G,H,I,J,K,L,M,N,O,P,Q,R,S,T,U,V,W,X,Y,Z}

\theoremstyle{plain}

\theoremstyle{definition}

\title{Efficient sequential Bayesian inference for state-space epidemic models using ensemble data assimilation}

\author{
  Dhorasso Temfack\,\orcidlink{0009-0007-0630-7504}\thanks{School of Computer Science and Statistics, Trinity College Dublin, Ireland. Email: \href{mailto:temfackd@tcd.ie}{temfackd@tcd.ie}} \and
 Jason Wyse\,\orcidlink{0000-0003-1391-7371}\footnotemark[1]
}

\date{} 

\begin{document}

\maketitle

\begin{abstract}
Estimating latent epidemic states and model parameters from partially observed, noisy data remains a major challenge in infectious disease modeling. State-space formulations provide a coherent probabilistic framework for such inference, yet fully Bayesian estimation is often computationally prohibitive because evaluating the observed-data likelihood requires integration over a latent trajectory. The Sequential Monte Carlo squared (SMC$^2$) algorithm offers a principled approach for joint state and parameter inference, combining an outer SMC sampler over parameters with an inner particle filter that estimates the likelihood up to the current time point. Despite its theoretical appeal, this nested particle filter imposes substantial computational cost, limiting routine use in near-real-time outbreak response. We propose Ensemble SMC$^2$ (eSMC$^2$), a computationally efficient variant that replaces the inner particle filter with an Ensemble Kalman Filter (EnKF) to approximate the incremental likelihood at each observation time. While this substitution introduces bias via a Gaussian approximation, we mitigate finite-sample effects using an unbiased Gaussian density estimator and adapt the EnKF for epidemic data through state-dependent observation variance. This makes our approach particularly suitable for overdispersed incidence data commonly encountered in infectious disease surveillance. Simulation experiments with known ground truth and an application to 2022 United States (U.S.) monkeypox incidence data demonstrate that eSMC$^2$ achieves substantial computational gains while producing posterior estimates comparable to SMC$^2$. The method accurately reconstructs epidemic trajectories and estimates key epidemiological parameters, providing an efficient framework for sequential Bayesian inference from imperfect surveillance data.
\end{abstract}

\noindent\textbf{Keywords:} Bayesian inference, Ensemble Kalman Filter, State-space models, SEIR model, Epidemics.

\section{Introduction}
Accurate real-time inference of epidemic dynamics is essential for situational awareness, forecasting, and evaluating interventions during outbreaks. Public health decisions rely on rapidly updating estimates of quantities such as the time-varying reproduction number, incubation period, and infectious duration. Epidemic surveillance data are often noisy, incomplete, and reported irregularly, making the recovery of latent infection processes a challenging inverse problem: the unobserved transmission process must be inferred from partial, indirect observations of incidence or prevalence \citep{golightly2023accelerating, whitaker2025sequential}. This motivates a state-space modeling framework, in which latent epidemic states evolve according to mechanistic transmission dynamics, such as the Susceptible–Exposed–Infectious–Removed (SEIR) model, while observations provide indirect information about the underlying epidemic process.

State-space models (SSMs) have become central to modern infectious disease modeling because they explicitly account for stochastic transmission and observational uncertainty \citep{Dureau2013, birrell2018evidence}. Within this framework, Bayesian inference provides a principled approach to jointly estimate latent epidemic trajectories and epidemiological parameters. However, fully Bayesian inference for SSMs remains computationally demanding: the observed-data likelihood is typically intractable, and standard Monte Carlo methods may be impractical for large-scale or real-time applications.

Particle filtering and related methods have made Bayesian state estimation tractable in epidemic modeling, with applications to influenza, Ebola, COVID-19, and more recently, monkeypox \citep{Dureau2013, funk2018real, ghosh2023approximate, papageorgiou2024novel}. Particle Markov Chain Monte Carlo (PMCMC) \citep{andrieu2010particle} enables exact Bayesian inference by combining MCMC with unbiased likelihood estimates from a particle filter (PF), but it can be computationally expensive since a full PF must be run at each MCMC iteration. Likelihood estimates may also have high variance, leading to poor mixing and increased Monte Carlo error, especially when the data provide limited information about parameters. Sequential Monte Carlo Squared (SMC$^2$) \citep{chopin2013smc2} extends PMCMC by maintaining a population of parameter particles, each with an internal PF for state inference and likelihood evaluation. Parameter particles are reweighted, and a resample–move step with a PMCMC mutation kernel mitigates degeneracy and maintains diversity. This nested structure allows sequential Bayesian updating as new data arrive and has been applied successfully in epidemic modeling \citep{golightly2018efficient, TEMFACK2025100847}, but computational cost grows with particle number and weight degeneracy remains a concern \citep{snyder2008obstacles}. Using the Auxiliary Particle Filter (APF) \citep{pitt1999filtering} within SMC$^2$ can improve particle selection and reduce the number of particles required \citep{golightly2018efficient}, but it adds computational complexity due to more elaborate weighting and proposal steps. Likelihood-free approaches, notably Approximate Bayesian Computation (ABC), offer an alternative when the likelihood is fully intractable \citep{li2025advances}. However, ABC typically requires many model simulations and careful choice of summary statistics and thresholds, which can make real-time application challenging.

As a computationally efficient and appealing method, the Ensemble Kalman Filter \citep[EnKF,][]{evensen1994sequential} has demonstrated strong performance in data assimilation for near-Gaussian systems. Compared with PF, the EnKF can achieve reliable results with substantially fewer particles (also referred to as ensemble members), since it updates the ensemble through linear transformations rather than importance resampling, thereby mitigating weight degeneracy. In practice, this often allows the EnKF to outperform particle filters in moderately nonlinear settings where the latter suffer from sample depletion \citep{morzfeld2017collapse}. In epidemiological applications, the EnKF is used to track infection dynamics and estimate transmission parameters \citep{mitchell2021analyzing, lal2021application, sun2023analysis}. Most implementations employ an augmented EnKF to jointly estimate states and parameters \citep{lal2021application, abbas2025joint}. While effective, this approach can be less flexible in adapting to changing transmission dynamics, since parameters are treated as additional static states and their artificial evolution (e.g., via random-walk perturbations) may lead to filter degeneracy or drift over time, a well-known issue shared by most Bayesian augmented filtering methods \citep{stroud2018bayesian}.

Several hybrid approaches have been developed to combine the scalability of the EnKF with the inferential rigor of full Bayesian methods. For instance, \cite{stroud2018bayesian} proposed a two-layer filter pairing an EnKF for states with a particle filter for parameters, while \cite{katzfuss2020ensemble} suggested replacing the particle filter with an EnKF within the particle MCMC framework, an approach that substantially reduces computational cost at the expense of a small approximation bias. Extending this idea, \cite{drovandi2022ensemble} introduced the Ensemble MCMC (eMCMC) algorithm, which corrects the bias in the EnKF likelihood using an unbiased Gaussian density estimator together with an early-rejection scheme to improve efficiency. Their results showed that EnKF-based likelihoods can achieve accurate posterior inference in moderately nonlinear systems with orders-of-magnitude speedups compared to particle filters. 

In this work, we propose Ensemble SMC$^2$ (eSMC$^2$), a novel algorithm that integrates the EnKF into the inner layer of the SMC$^2$ framework to enable fully sequential Bayesian inference over both latent states and parameters. This approach substantially reduces computational cost while maintaining accurate posterior inference. To mitigate the bias inherent in the standard EnKF likelihood estimate, we incorporate the unbiased Gaussian likelihood correction employed by \cite{drovandi2022ensemble}, which reduces sensitivity to the number of ensemble members. While we cannot provide formal bias quantification, prior studies indicate that EnKF-based likelihoods yield stable and accurate inference in moderately nonlinear systems \citep{khalil2015estimation, katzfuss2020ensemble, drovandi2022ensemble}. We further employ a variant inspired by the Poisson Kalman Filter \citep{ebeigbe2020poisson}, making the approach suitable for observations typical in disease surveillance and relaxing the Gaussian noise assumptions of standard EnKF methods. We demonstrate its performance on a diffusion-driven SEIR model with incidence data aggregated over fixed reporting intervals, showing that eSMC$^2$ delivers an accurate estimate of the time-varying transmission dynamics and key epidemiological parameters using both synthetic and real datasets.

In parallel to our work, \cite{golightly2025nested} independently proposed an extension of SMC2 in which the inner particle filter is replaced by an EnKF. Their nested EnKF (NEnKF) maintains parameter diversity via a resample–move strategy, combined with delayed-acceptance steps to improve efficiency. While both approaches embed an EnKF within SMC2, they differ in their motivating application and methodological focus. \cite{golightly2025nested} consider a broad class of state-space models, whereas our method is motivated by epidemic inference and introduces an unbiased Gaussian likelihood correction with variance adaptation for count-valued data. Nevertheless, both approaches highlight the potential of embedding the EnKF within SMC2 as an efficient alternative to nested particle filter for sequential state and parameter inference.

The remainder of this paper is organized as follows. In Section \ref{sec: ssm}, we present the SEIR model within a state-space framework, review the EnKF, and describe the eSMC$^2$ algorithm. Section \ref{sec:result} presents simulation studies and an application to U.S. monkeypox incidence data. Finally, Section \ref{sec: discussion} discusses the limitations of our approach, potential improvements, and directions for future research.
\section{Methodology}\label{sec: ssm}

\subsection{SEIR model in a Bayesian state-space framework} \label{sec:ssm}

We model the spread of infectious diseases using a stochastic SEIR framework that explicitly tracks both latent epidemic states and the time-varying transmission rate. While we present the SEIR model here as a concrete example, the methodology is applicable to a wide range of compartmental epidemic models. Let the latent state at time $t$ be $x_t = (S_t, E_t, I_t, R_t, Z_t, \log (\beta_t))^{\top}$, where $S_t, E_t, I_t, R_t$ denote the numbers of susceptible, exposed (infected but not yet infectious), infectious, and removed (recovered or deceased) individuals, respectively. The variable $Z_t$ is not a dynamical compartment but an auxiliary quantity representing the new incidence over each reporting interval, included for convenience to directly match the observation model. The variable $\log (\beta_t)$ is incorporated into the latent state to allow inference on the time-varying transmission rate. The latent dynamics follow the system
\begin{align}\label{eq:seir}
\begin{cases}
\displaystyle
\frac{dS_t}{dt} = -\beta_t \frac{S_t I_t}{N}, \quad
\frac{dE_t}{dt} = \beta_t \frac{S_t I_t}{N} - \alpha E_t, \quad
\frac{dI_t}{dt} = \alpha E_t - \gamma I_t, \quad
\frac{dR_t}{dt} = \gamma I_t, \\[2mm]
\displaystyle
Z_t = \int_{t-1}^{t} \alpha E_s \, ds, \\[2mm]
\displaystyle
d\log(\beta_t) = \nu_\beta \, dB_t.
\end{cases}
\end{align}
where $N = S_t + E_t + I_t + R_t$ is the constant population size, and $1/\alpha$ and $1/\gamma$ denote the average incubation and infectious periods, respectively. The transmission rate $\beta_t$ is modeled as a stochastic process on the log scale, following a Brownian motion with volatility $\nu_\beta$ and driven by a standard Wiener process $B_t$. This allows $\beta_t$ to vary over time in response to behavioral changes, public health interventions, or seasonal effects \citep{Dureau2013, cazelles2018accounting}. The parameter $\nu_\beta$ controls the magnitude of these fluctuations: larger values permit faster changes, while smaller values constrain variation. This formulation enables the model to capture unobserved temporal variability in transmission intensity throughout an epidemic. The effective reproduction number is defined as $R_{\text{eff}}(t) = \beta_t S_t / (\gamma N)$. To simulate sample trajectories, the continuous-time system \eqref{eq:seir} is propagated forward using a discrete-time approximation. Let $\theta$ denote the vector of unknown model parameters governing both the latent dynamics in \eqref{eq:seir} and the observation model described below. The latent state is updated according to
\begin{align}\label{state_process}
    x_t|x_{t-1}, \theta \sim p(.| x_{t-1}, \theta).
\end{align}
The forward process $p(.|x_{t-1}, \theta)$ updates $S_t, E_t, I_t, R_t,$ and $Z_t$ deterministically via a forward Euler scheme, while $\log (\beta_t)$ evolves stochastically using the Euler–Maruyama method \citep{kloeden2011stochastic}. Formally, the deterministic updates are represented as a Dirac delta distribution within the transition density $p(x_t|x_{t-1}, \theta)$, whereas the stochasticity of $\log (\beta_t)$ introduces process noise.  Details on the discretization and transition density are provided in Appendix \ref{appA}.

We assume that the available data $y_{1:t} = \{y_1, \ldots, y_t\}$, representing daily reported cases, are noisy measurements of the latent incidence $Z_t$, linked through the conditional distribution $p(y_t | x_t, \theta) = p(y_t | Z_t, \theta)$. A natural starting point is to model the counts as realizations of a Poisson random variable. In practice, however, daily case counts often fluctuate more than a Poisson model would predict, a phenomenon known as overdispersion. To account for this, we also consider a Negative Binomial formulation.

Specifically, the number of reported cases $y_t$ in a reporting interval $(t-1, t]$, $t = 1, \dots, T$, is modeled as either
\begin{subequations}\label{obs_dis}
\begin{align} 
&y_t | x_t, \theta \sim \text{Poisson}(\rho Z_t) \quad \text{or}  \\
&y_t | x_t, \theta \sim \text{NegBin}\big(\mathrm{mean} = \rho Z_t, ~\mathrm{var} = \rho Z_t + \phi (\rho Z_t)^2 \big),
\end{align}
\end{subequations}
The Negative Binomial is parameterized by its mean $\rho Z_t$ and variance $V_t=\mathrm{Var}[y_t|x_t] =\rho Z_t + \phi(\rho Z_t)^2$, where $\phi > 0$ is the overdispersion parameter; the Poisson model, with $V_t=\mathrm{Var}[y_t|x_t] = \rho Z_t$, is recovered as $\phi \to 0$. This mean-variance parameterization is used directly in the ensemble-based observation variance described in Section \ref{sec_poiss} below, ensuring consistency between the observation model and the EnKF update step. $\rho \in (0,1]$ is the reporting fraction, informed by expert knowledge. Specifying $\rho$ helps mitigate practical non-identifiability issues highlighted in \cite{cazelles2018accounting}, since decreases in observed cases could otherwise be attributed equally to a lower reporting rate, a reduced transmission rate, or a combination of both. Practical considerations regarding the structural identifiability of parameters in such models were also discussed in \cite{TEMFACK2025100847}. 

Together, Eqs~\eqref{state_process}--\eqref{obs_dis} define our complete state-space model \citep{birrell2018evidence}. Let $x_{0:t}=\{x_0,\ldots,x_t\}$ denote the latent states up to time $t$. Prior knowledge about $\theta$ and initial state $x_0$ is encoded in the distributions $p(\theta)$ and $p(x_0|\theta)$, respectively. The joint posterior of states and parameters given the observations is then
\begin{align}
p(x_{0:t}, \theta | y_{1:t}) \propto p(\theta)\, p(x_0|\theta)\, \prod_{s=1}^{t} \Big[p(x_s | x_{s-1}, \theta)\, p(y_s | x_s, \theta)\Big].
\end{align}
In this sequential learning setting, the inferential objectives at each time $t$ are twofold. First, for a given fixed value of $\theta$, filtering aims to infer the current state $x_t$ conditional on all observations up to that time, corresponding to:
\begin{align}\label{pb_x}
p(x_t | y_{1:t}, \theta) \propto p(y_t | x_t, \theta)\underbrace{\int p(x_t | x_{t-1}, \theta)\, p(x_{t-1} | y_{1:t-1}, \theta)\, dx_{t-1}}_{\text{Forecast density } p(x_{t}|y_{1:t-1}, \theta)}.
\end{align}
Second, parameter estimation focuses on the marginal posterior of $\theta$, 
\begin{align}\label{pb_theta}
p(\theta|y_{1:t}) &\propto p(\theta)\, \prod_{s=1}^{t} p(y_s | y_{1:s-1}, \theta) \notag\\
&\propto p(\theta) \underbrace{\int \prod_{s=1}^{t} \Big[p(y_s | x_s, \theta)\, p(x_s | y_{1:s-1}, \theta)\Big]\, dx_s}_{\text{marginal likelihood } p(y_{1:t} | \theta)} ,
\end{align}
where we adopt the convention that $p(y_1|\theta) = p(y_1 | y_{1:0},\theta)$. The filtering distribution that accounts for parameter uncertainty is obtained by marginalising the conditional filtering distribution over the parameter posterior, i.e.
\begin{align}
p(x_t | y_{1:t}) = \int p(x_t | y_{1:t}, \theta)\, p(\theta | y_{1:t})\, d\theta.
\end{align}
For nonlinear state-space models, Eqs~\eqref{pb_x} and~\eqref{pb_theta} cannot be solved analytically and are generally intractable. Consequently, practical inference relies on numerical approximations such as particle filtering or EnKF-based methods to generate samples from the filtering distribution or to estimate the marginal likelihood. In this work, motivated by the computational efficiency and scalability considerations discussed in the introduction, we focus on EnKF-based inference.

\subsection{Ensemble Kalman Filter}\label{sec_poiss}
We now describe the ensemble Kalman filter \citep[EnKF,][]{evensen1994sequential}, a sequential inference method for state-space models with nonlinear latent dynamics. While the EnKF was originally developed for systems with Gaussian observations, many real-world datasets, including those arising in epidemic surveillance often deviate from this assumption and may exhibit overdispersion. To accommodate such data, we follow the approach of \cite{ebeigbe2020poisson} and extend it by constructing observation noise terms consistent with the observation models in Eq \eqref{obs_dis}.

The EnKF alternates between prediction and update operations at each time step to sequentially refine the latent state estimate based on incoming observations (Fig~\ref{Fig1}). During the prediction step, the ensemble is propagated forward through the system dynamics, while in the update step, the ensemble is adjusted using the most recent observation.

\begin{figure}[H]
    \centering
    \definecolor{darkviolet}{RGB}{108,56,130}
    \definecolor{lightviolet}{RGB}{219,100,187}

    \begin{tikzpicture}[scale=0.55, every node/.style={transform shape}, >=latex, thick]
        \tikzstyle{update} = [rectangle, rounded corners=2pt, minimum width=3cm, minimum height=1.3cm, 
                              text centered, draw=darkviolet, fill=darkviolet, text=white, font=\Large\bfseries]
        \tikzstyle{prediction} = [rectangle, rounded corners=2pt, minimum width=3cm, minimum height=1.3cm, 
                                  text centered, draw=lightviolet, fill=lightviolet, text=white, font=\Large\bfseries]

        \node[draw=none, fill=none] (dot1)  at (-3,0) {$\bm{\cdots}$}; 
        \node[update] (update0) at (0,0) {Update};
        \node[prediction] (pred0) at (5,0) {Prediction};
        \node[update] (update1) at (10,0) {Update};
        \node[prediction] (pred1) at (15,0) {Prediction};
        \node[update] (update2) at (20,0) {Update};
        \node[draw=none, fill=none] (dot2)  at (23,0) {$\bm{\cdots}$};

        \draw[->,thick, >=latex] (update0) -- (pred0);
        \draw[->,thick, >=latex] (pred0) -- (update1);
        \draw[->,thick, >=latex] (update1) -- (pred1);
        \draw[->,thick, >=latex] (pred1) -- (update2);

        \draw[->,thick, >=latex] (-2.3,0) -- (update0);
        \draw[->,thick, >=latex] (update2) -- (22.3,0);

        \node[font=\Large] at (0,2.1) {$\bm{y_{t-1}}$};
        \draw[->,thick, >=latex] (0,1.8) -- (0,0.7);

        \node[font=\Large] at (10,2.1) {$\bm{y_t}$};
        \draw[->,thick, >=latex] (10,1.8) -- (10,0.7);

        \node[font=\Large] at (20,2.1) {$\bm{y_{t+1}}$};
        \draw[->,thick, >=latex] (20,1.8) -- (20,0.7);

        \node[font=\large] at (2.5,-1.0) {$\bm{p(x_{t-1}|y_{1:t-1}, \theta)}$};
        \node[font=\large] at (7.5,-1.0) {$\bm{p(x_{t}|y_{1:t-1}, \theta)}$};
        \node[font=\large] at (12.5,-1.0) {$\bm{p(x_{t}|y_{1:t}, \theta)}$};
        \node[font=\large] at (17.5,-1.0) {$\bm{p(x_{t+1}|y_{1:t}, \theta)}$};

        \draw[dashed, line width=0.4pt] (0,-1.4) -- (0,-2.1);
        \draw[dashed, line width=0.4pt] (10,-1.4) -- (10,-2.1);
        \draw[dashed, line width=0.4pt] (20,-1.4) -- (20,-2.1);

        \node[font=\normalsize, align=center] at (0,-2.3) {$\bm{t-1}$};
        \node[font=\normalsize, align=center] at (10,-2.3) {$\bm{t}$};
        \node[font=\normalsize, align=center] at (20,-2.3) {$\bm{t+1}$};

        \draw[->,thick, >=latex, line width=0.4pt] (-3,-2.7) -- (23,-2.7) node[pos=0.96, above] {Time};
    \end{tikzpicture}

    \caption{\footnotesize \textbf{Schematic illustration of the sequential update–prediction cycle in the EnKF}. 
    Each time step alternates between an update stage, where observations $y_t$ are assimilated to refine the latent state estimate, and a prediction stage, where the ensemble is propagated forward through the system dynamics.}
    \label{Fig1}
\end{figure}

Let us first suppose that the parameters $\theta$ and a sequence of observations $y_{1:T}$ (in our case, the daily incidence) are known. At each time step $t$, the EnKF aims to infer the latent state $x_t$ given the data $y_{1:t}$. Let the ensemble at time $t-1$ be denoted by $\{x_{t-1}^{(i)}\}_{i=1}^{N_{x}}$, representing samples from the conditional $p(x_{t-1} | y_{1:t-1}, \theta)$. The forecast ensemble $\{x_t^{(f,i)}\}_{i=1}^{N_{x}}$ is generated by propagating each particle forward through the transition dynamics, that is,
\begin{align}
  x_t^{(f,i)} \sim p(. | x_{t-1}^{(i)}, \theta), \quad i = 1, \ldots, N_{x}. 
\end{align}
The forecast density $p(x_t | y_{1: t-1})$ is then approximated by:
\begin{align}\label{forcas_dist}
p(x_t | y_{1:t-1}, \theta) \approx \mathcal{N}\big(x_t;~\widehat{\mu}_{t|t-1}, ~\widehat{\Sigma}_{t|t-1}\big),
\end{align}
where $\widehat{\mu}_{t|t-1}$ and $\widehat{\Sigma}_{t|t-1}$ are the sample mean and covariance of the forecast ensemble.

A key adaptation for epidemic data is the specification of the observation error variance $V_t$. Classical Kalman filtering assumes Gaussian observations with constant variance  (or at least does not depend on the latent state), an assumption often violated for epidemiological models, where the variability of reported incidence typically scales with its magnitude \citep{ebeigbe2020poisson}. A principled approach is to set $V_t$ to match the expected conditional variance of the observation given the latent state $\mathrm{E}\big[\mathrm{Var}[y_t|x_t]\big]$, as implied by the observation model \eqref{obs_dis}. In practice, we approximate the expectation above using the forecast ensemble.  We define the ensemble estimate of the observation error  variance by
\begin{align}\label{V_est}
\mathrm{E}\big[\mathrm{Var}[y_t|x_t]\big] \approx V^{N_x}_t&= \frac{1}{N_x} \sum_{i=1}^{N_x} \mathrm{Var}\big[y_t|x_t^{(f,i)}\big],  \notag\\[4pt] 
&=\begin{cases}
\frac{1}{N_x} \sum_{i=1}^{N_x}Hx_t^{(f,i)} , & \text{Poisson model}, \\[5pt] 
\frac{1}{N_x} \sum_{i=1}^{N_x}\big[Hx_t^{(f,i)} + \phi\big(Hx_t^{(f,i)}\big)^2\big], & \text{Negative Binomial model}, 
\end{cases} 
\end{align}
where $H=[0,0,0,0,\rho, 0]$ projects the model state onto the observed component, i.e., $Hx_t^{(f,i)}=\rho Z_t^{(f,i)}$, $i=1,\cdots, N_x$. To prevent numerical instability when $V^{N_x}_t$ is very small (or near zero), we regularize the variance estimate by setting
\begin{align}
\widehat{V}_t=\max\big(\eta, V_t^{N_x}\big),
\end{align} 
where in our implementation we set $\eta = 0.1$, small relative to the typical magnitude of $V_t^{N_x}$. We found that results are insensitive to this choice provided $\eta < 1$. Appendix \ref{appA} provides a detailed justification for this specification.  This specification allows the EnKF update to capture key features of the observation distribution \eqref{obs_dis}, though it can overestimate uncertainty when case counts are extremely low. Such low-incidence scenarios are not the primary focus of our study, and alternative sequential methods, such as the Lifebelt particle filter \citep{corbella2024lifebelt}, may be preferred in those cases. A similar treatment of adaptive variance specification is discussed in (Section 3.2, \cite{asfaw2024tutorial}).

During the analysis (update) step, each forecast ensemble member is adjusted using the Kalman gain and the innovation. In the stochastic (perturbed-observation) EnKF, the update is
\begin{align}
x_t^{(i)} = x_t^{(f,i)} + K_t \left( y_t + v_t^{(i)} - H x_t^{(f,i)} \right), \quad v_t^{(i)} \sim \mathcal{N}(0, \widehat{V}_t),
\end{align}
where the Kalman gain is given by
\begin{align}
K_t = \widehat{\Sigma}_{t|t-1} H^\top \left( H \widehat{\Sigma}_{t|t-1} H^\top + \widehat{V}_t, \right)^{-1}.
\end{align}

Using the Gaussian approximation of the forecast density in \eqref{forcas_dist}, the one-step predictive likelihood can be expressed as
\begin{align}\label{pred_lik}
p(y_t | y_{1:t-1}, \theta) \approx \mathcal{N}\Big(y_t;~H \widehat{\mu}_{t|t-1},~H \widehat{\Sigma}_{t|t-1} H^\top + \widehat{V}_t\Big).
\end{align}
Only the key components of the EnKF algorithm for state inference are presented above; for a comprehensive overview, see \cite{katzfuss2020ensemble}.  For linear–Gaussian SSMs, the likelihood approximation in \eqref{pred_lik}  converges to the true likelihood as $N_x \to \infty$, but for finite ensembles it remains biased due to sampling variability in the forecast mean and covariance \citep{drovandi2022ensemble}. Importantly, even if the forecast ensemble is exactly Gaussian, the estimated density $\mathcal{N}(y_t; ~H \widehat{\mu}_{t|t-1}, ~H \widehat{\Sigma}_{t|t-1} H^\top +  \widehat{V}_t)$ is not an unbiased estimate of the true likelihood. This occurs because the empirical normal density $\mathcal{N}(y;~ \mu_{_{N_x}}, \Sigma_{_{N_x}})$, computed from the finite-ensemble mean $\mu_{_{N_x}}$ and covariance $\Sigma_{_{N_x}}$, is biased estimator of the true density $\mathcal{N}(y; \mu, \Sigma)$ for any finite $N_x$.

Nevertheless, empirical results in the literature \citep{khalil2015estimation,  stroud2018bayesian, drovandi2022ensemble} show that the EnKF likelihood performs remarkably well for inference in both linear and nonlinear systems, with posterior estimates only weakly dependent on the ensemble size \citep{drovandi2022ensemble}. To mitigate the finite-sample bias, we adopt the unbiased Gaussian density estimator of \cite{ghurye1969unbiased}, later employed by \cite{price2018bayesian} and  \cite{drovandi2022ensemble}. This estimator yields an unbiased estimate of $\mathcal{N}(y;\mu, \Sigma)$ provided $N_x > d_y + 3$, where $d_y$ is the dimension of $y$. It is defined as
\begin{align}
\widehat{\mathcal{N}}_{\text{unb}}(y;~ \mu_{_{N_x}}, \Sigma_{_{N_x}}) = &\ (2\pi)^{-d_y/2} \frac{c(d_y, N_x-2)}{c(d_y, N_x-1)(1 - 1/N_x)^{d_y/2}} |M_{_{N_x}}|^{-(N_x-d_y-2)/2} \notag \\
& \times \psi\left(M_{_{N_x}} - \frac{(y - \mu_{_{N_x}})(y - \mu_{_{N_x}})^\top}{1 - 1/N_x} \right)^{(N_x-d_y-3)/2},
\end{align}
where $M_{_{_{N_x}}} = (N_x - 1)\Sigma_{_{N_x}}$, and $\psi(A) = |A|$ if $A$ is positive definite, and $0$ otherwise. Here, $|A|$ denotes the determinant of the matrix $A$ and the constants $c(d, v)$ are defined as
\begin{align}
   c(d_y, v) = \dfrac{2^{-d_yv/2} \pi^{-d_y(d_y-1)/4}}{\prod_{i=1}^{d_y} \Gamma\left(\frac{1}{2}(v - i + 1)\right)}.
\end{align}
We can now replace the standard EnKF likelihood term in \eqref{pred_lik}  with its unbiased counterpart:
\begin{align}
        \widehat{p}_{\text{enkf}}^{N_x}(y_t | y_{1:t-1},\theta)= \widehat{\mathcal{N}}_{\text{unb}}\left(y_t ; ~ H \widehat{\mu}_{t|t-1}, ~H \widehat{\Sigma}_{t|t-1} H^\top + \widehat{V}_t\right).
\end{align}
The  approximation of the  marginal likelihood up to the time  $t$ is then  given by:
\begin{align}\label{lik_est}
   \widehat{p}_{\text{enkf}}^{N_x}(y_{1:t} | \theta) = \prod_{s=1}^{t} \widehat{p}_{\text{enkf}}^{N_x}(y_s | y_{1:s-1}, \theta).
\end{align}
Unlike in the case of the particle filter \citep{andrieu2010particle},  the resulting posterior approximation  $\widehat{p}_{\text{enkf}}^{N_x}(\theta|y_{1:t}) \propto\widehat{p}_{\text{enkf}}^{N_x}(y_{1:t} |\theta)\,p(\theta)$, still does not target the true posterior (due to the Gaussian assumption). However, when the Gaussian assumption holds, the resulting posterior no longer depends on  $N_x$, eliminating the finite-ensemble bias and yielding a stable target for Bayesian inference \citep{ghurye1969unbiased, drovandi2022ensemble}. The EnKF procedure, including unbiased likelihood estimation, is outlined in Algorithm~\ref{alg:pois_enkf}. This estimator can then be incorporated into a Metropolis–Hastings algorithm to target the sequence of posterior distributions $\{p(\theta),~ p(\theta | y_{1}),~p(\theta |y_{1:2}),\ldots,~p(\theta | y_{1:T})\}$.

\begin{algorithm}[H]
\caption{Ensemble Kalman filter}\label{alg:pois_enkf}
Operations involving index $i$ must be performed for $i = 1,..., N_x$.

\textbf{Inputs:} Observation $y_{1:T}$, Number of ensemble members $N_x$, Parameter vector $\theta$.

\textbf{Output:} Filtering ensembles $\{x_{0:T}^{(i)}\}_{i=1}^{N_{x}}$ and likelihood estimate $\widehat{p}_{\text{enkf}}^{N_x}(y_{1:T} | \theta)$

\hrulefill 
\begin{algorithmic}[1]
    \State Sample initial  particles : $x^{(i)}_0 \sim p( x_0)$
    \For{$t=1$ to $T$}
        \State  Sample forecast ensemble as: $x_t^{(f,i)} \sim p( .|x_{t-1}^{(i)},\theta)$
        \State  Compute the observation variance, ensemble forecast mean, and variance: $ \widehat{V}_t$, $\widehat{\mu}_{t|t-1}$, and $\widehat{\Sigma}_{t|t-1}$.
        \State Compute the incremental Likelihood:
            \[
        \widehat{p}_{\text{enkf}}^{N_x}(y_t | y_{1:t-1},\theta)= \widehat{\mathcal{N}}_{\text{unb}}\left(y_t ; ~ H \widehat{\mu}_{t|t-1}, ~H \widehat{\Sigma}_{t|t-1} H^\top +  \widehat{V}_t\right).
    \]
        \State   Shift each ensemble member as: 
        \[x_t^{(i)} = x_t^{(f,i)} + K_t (y_t + v_t^{(i)}- H x_t^{(f,i)}), \quad v_t^{(i)} \sim \mathcal{N}(0, \widehat{V}_t) \]
    \EndFor
\State Compute overall likelihood: $   \widehat{p}_{\text{enkf}}^{N_x}(y_{1:T} |, \theta) = \prod_{t=1}^{T} \widehat{p}_{\text{enkf}}^{N_x}(y_t | y_{1:t-1}, \theta).$
\end{algorithmic}
\end{algorithm}

\subsection{Ensemble SMC$^2$}

The EnKF algorithm described previously assumes that the model parameters $\theta$ are known. In practice, these parameters are rarely known, and it is preferable to assign a distribution $p(\theta)$ that reflects prior knowledge. To account for this uncertainty, inference on $\theta$ can be performed by embedding the EnKF within an SMC sampler over the parameter space \citep{del2006sequential}. This approach follows the general principle of SMC$^2$ \citep{chopin2013smc2}, which traditionally employs a particle filter to estimate the likelihood. At each observation time $t$, the posterior distribution of the parameters given data $y_{1:t}$ satisfies the recursive relation
\begin{align}
p(\theta | y_{1:t}) = \dfrac{p(y_t | y_{1:t-1}, \theta) \, p(\theta | y_{1:t-1})}{\int p(y_t |y_{1:t-1}, \theta) \, p(\theta |y_{1:t-1}) \, d\theta}.
\end{align}
A practical way to approximate this posterior is to represent it with a finite set of weighted parameter particles. Let assume that $\{\theta^{m}, \omega^{m}_{t-1}\}_{m=1}^{N_\theta}$ is particle approximation from the previous posterior $p(\theta | y_{1:t-1})$. The updated posterior can then be expressed as:
\begin{align}
p(\theta | y_{1:t}) \approx \sum_{m=1}^{N_\theta}\omega_{t}^{m} \, \delta_{\theta^{m}}(\theta),
\end{align}
where $\omega_{t}^{m}$, for $m=1,\ldots, N_{\theta}$ are the normalized weights given by
\begin{align}
\omega_t^{m} =  \dfrac{ \omega_{t-1}^{m}p(y_t | y_{1:t-1}, \theta^{m})}{\sum_{j=1}^{N_\theta}\omega_{t-1}^{j} p(y_t | y_{1:t-1}, \theta^{j})}.
\end{align}
As the incremental likelihood $p(y_t | y_{1:t-1}, \theta)$ is intractable  in a nonlinear system, at time $t$, the weight of the $\theta$-particle $\theta^m$ is updated using an estimator $\widehat{p}_{\text{enkf}}^{N_x}(y_t | y_{1:t-1}, \theta^m)$ obtained from
the associated EnKF. This differs from related work of \cite{wu2022ensemble}, where the EnKF is used to construct proposal kernels in the SMC sampler but not as a likelihood estimator. As in standard SMC$^2$, particle degeneracy may occur over time. To mitigate this, resampling (described in Appendix \ref{appB}) is triggered whenever the effective sample size (ESS) drops below $N_{\theta}/2$ \citep{chopin2013smc2, chopin2020introduction}. This is followed by a jittering mechanism that rejuvenates the particles by proposing new candidates, which are then accepted or rejected through a few iterations of Particle Marginal Metropolis--Hastings (PMMH). Since the weighted particles already approximate $ p(\theta| y_{1:t})$ (under the Gaussian assumption), applying a Metropolis--Hastings update to each one can only enhance this approximation (see Algorithm~\ref{alg:PMCMC}).

\begin{algorithm}[H]
\caption{PMMH mutation}\label{alg:PMCMC}
\textbf{Inputs:} Observations $y_{1:t}$, Current parameter $\theta$,  current likelihood
estimate: $\widehat{p}_{\text{enkf}}^{N_x}(y_{1: t} | \theta)$, Number of ensemble members $N_x$,  Number of PMMH moves $R$  \\
\textbf{Output:} New parameter $\theta'$, New likelihood: $\widehat{p}_{\text{enkf}}^{N_x}(y_{1: t} | \theta')$

\hrulefill
\begin{algorithmic}[1]
\For{$r=1$ to $R$}
    \State Propose $\theta^* \sim q(\theta^*|\theta)$
    \State Run a new EnKF (Algorithm~\ref{alg:pois_enkf}) with $\theta^*$ and compute $\widehat{p}_{\text{enkf}}^{N_x}(y_{1:t}|\theta^*)$
    \State Calculate acceptance ratio:
    \[
   \vartheta(\theta^{*}, \theta)=\min \left\{1,~ \dfrac{ \widehat{p}_{\text{enkf}}^{N_x}(y_{1: t} | \theta^*)p(\theta^*)}{\widehat{p}_{\text{enkf}}^{N_x}(y_{1: t} | \theta)p(\theta)}\times \dfrac{q(\theta|\theta^*)}{q(\theta^*|\theta)} \right\}
    \]
    \State Accept $\theta^*$ with probability $\vartheta(\theta^{*}, \theta)$, otherwise retain $\theta$
\EndFor
\end{algorithmic}
\end{algorithm}

In the numerical experiments presented in Section~\ref{sec:result}, we use an independent proposal distribution, $q(\theta^{*} | \theta) = q(\theta^{*})$. The proposal is constructed directly on the parameter space using a multivariate normal distribution. Specifically, new parameter particles $\theta^*$ are drawn from
\begin{align}
q(\theta^{*}) = \mathcal{N}\!\left(\theta^{*};\, \widehat{\mathrm{E}}[\theta | y_{1:t}], \, c\,\widehat{\mathrm{Var}}[\theta | y_{1:t}]\right),
\end{align}
where $\widehat{\mathrm{E}}[\theta | y_{1:t}]$ and $\widehat{\mathrm{Var}}[\theta | y_{1:t}]$ denote the empirical mean and covariance of the current particle population $\{\theta^m, \omega_t\}_{m=1}^{N_\theta}$, and $c$ is a scaling constant that controls the covariance matrix.

\begin{figure}[H]
    \centering
    \includegraphics[width=0.9\linewidth]{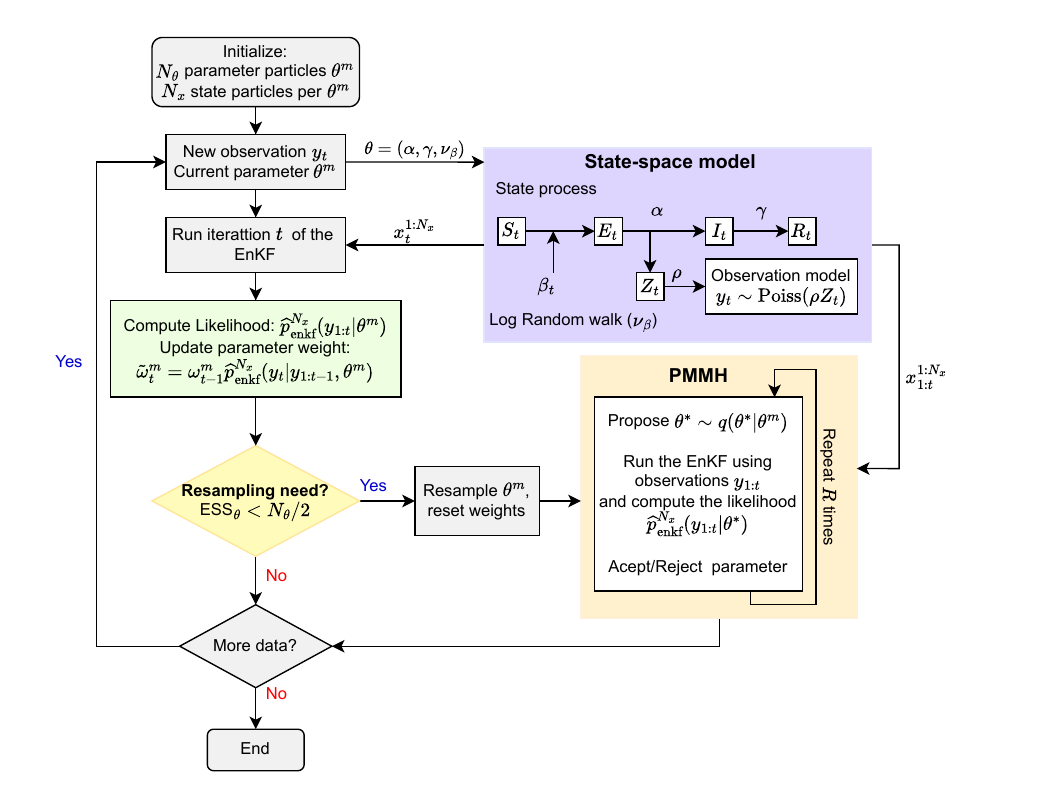}
\caption{\footnotesize 
\textbf{Flowchart of the eSMC$^2$ algorithm.} Each parameter particle carries an ensemble of state particles, which are propagated using the EnKF. Weights are updated based on an EnKF-based likelihood approximation. When the ESS falls below a threshold, parameter particles are resampled and rejuvenated via a PMMH step. This procedure is repeated for every data point, allowing sequential Bayesian learning.
}\label{Fig2}
\end{figure}

The eSMC$^2$ algorithm can be viewed as a variant of the standard SMC$^2$, where the particle filter–based likelihood is replaced by an EnKF approximation in \eqref{lik_est}. From a practical perspective, this modification is expected to have only a slight effect on the posterior estimates of $\theta$ and  $x_{1:t}$. The complete procedure is detailed in Algorithm~\ref{alg:esmc2}, with a schematic overview in Fig~\ref{Fig2}.

To propagate parameter uncertainty in the hidden state estimation, we approximate the marginal posterior distribution of the latent states at time $t$, $p(x_t| y_{1:t})$, by integrating over a subset of parameter particles, $\{\theta^{j}\}_{j=1}^{N_c} \sim p(\theta | y_{1:t})$, drawn from Algorithm \ref{alg:esmc2} output \citep{steyn2025primer}. In this work, we set 
$N_c=100$, which in our numerical examples (Section \ref{sec:result}) provides satisfactory coverage of the posterior. For each sampled $\theta^{j}$, an EnKF with $N_x$ ensemble members is run to approximate the filtering distribution $p(x_t | y_{1:t}, \theta^{j})$, and the resulting trajectories are combined across all parameter samples to form an empirical approximation of the  state posterior:
\begin{align}
    p(x_t | y_{1:t}) \approx \frac{1}{N_c} \sum_{j=1}^{N_c} p(x_t| y_{1:t}, \theta^{j}),
\end{align}
resulting in a total of $N_c N_x$ trajectories that approximate the filtering distribution.

\begin{algorithm}[H]
\caption{Ensemble Sequential Monte Carlo Squared (eSMC$^2$)}\label{alg:esmc2}
Operations involving index $m$ must be performed for $m = 1,..., N_{\theta}$.

\textbf{Inputs:} Observations $y_{1:T}$, prior $p(\theta)$, Number of ensemble members $N_x$, Number of parameter particles: $N_{\theta}$, Number of PMMH moves $R$ \\
\textbf{Output:} Parameter particles:  $\left\{\theta^m_{0:t}\right\}_{m=1}^{N_{\theta}}$

\hrulefill
\begin{algorithmic}[1]
\State Initialize $\theta_0^m \sim p(\theta)$, weights $\omega_0^m = 1/N_\theta$, for $m=1,\ldots,N_\theta$
\For{$t=1$ to $T$}
    \State Perform iteration $t$ of EnKF (Algorithm~\ref{alg:pois_enkf}) to estimate $\widehat{p}_{\text{enkf}}^{N_x}(y_t | y_{1:t-1}, \theta_t^m)$
    \State Update weights: 
    \[\tilde{\omega}_t^m = \omega_{t-1}^m \times \widehat{p}_{\text{enkf}}^{N_x}(y_t | y_{1:t-1}, \theta_t^m)\]

    \State Normalize weights: $\omega_t^m = \tilde{\omega}_t^m / \sum_{j=1}^{N_\theta} \tilde{\omega}_t^j$ and compute ESS$_{\theta}=1/\sum_{m}^{N_\theta}(\omega_t^m)^2$
    \If{ESS$_{\theta} < N_\theta$/2}
        \State Resample parameters $\{\theta_t^m\}$ with replacement according to $\{\omega_t^m\}$
        \State Reset weights $\omega_t^m = 1/N_\theta$
        \State Perform PMMH move (Algorithm~\ref{alg:PMCMC}) for each $\theta_t^m$ 
    \EndIf
\EndFor
\end{algorithmic}
\end{algorithm}

It is important to note a key theoretical distinction between SMC$^2$ and eSMC$^2$. Standard SMC$^2$ is ``exact'': as the number of state and parameter particles $N_x, N_\theta \to \infty$, it converges to the true posterior $p(\theta|y_{1:t})$. In contrast, eSMC$^2$ is generally biased because the EnKF approximates the incremental likelihood $p(y_t|y_{1:t-1},\theta)$ via a Gaussian assumption. Nevertheless, the EnKF likelihood typically has low variance \citep{katzfuss2020ensemble}, and unbiased corrections are possible under mild conditions \citep{drovandi2022ensemble}. Empirical studies show that EnKF-based likelihoods yield accurate and stable inference for both linear and moderately nonlinear SSMs \citep{khalil2015estimation, drovandi2022ensemble}. Thus, while not formally exact, eSMC$^2$ provides practically accurate posterior estimates at much lower computational cost.

\section{Results}\label{sec:result}
\subsection{Inference on a simulated epidemic} \label{sec_sim}

To assess the performance of the proposed eSMC$^2$ algorithm, we first conducted hree simulation studies using the SEIR model from Section~\ref{sec: ssm}.  All algorithms were implemented in Python and executed on a desktop computer equipped with an Intel Core i7-1300H processor (3.40 GHz). Unless otherwise stated, simulations assumed a population size of $N = 500{,}000$ with initial conditions $I_0 = 10$, $S_0 = N-I_0$, and $E_0 = R_0 = 0$. Daily incidence $y_t$ was generated according to the Poisson observation model in Eq~\eqref{obs_dis}, assuming perfect case reporting ($\rho = 1$).

The three examples differ primarily in the temporal structure of the transmission rate $\beta_t$ and the number of observation points. Example~1 depicts a moderately time-varying epidemic with an initial surge followed by a mild resurgence, capturing short-term fluctuations in transmission. Example~2, by contrast, exhibits a smoother and more prolonged pattern that sustains infections for longer and decays more gradually. The latter setting was designed to emulate the dynamic observed in the 2022 U.S. mpox epidemic, where behavioral changes and intermittent control measures led to prolonged and uneven declines in case counts. Example~3 is motivated by epidemics such as COVID-19 or seasonal influenza that exhibit secondary infection waves, arising from imperfect immunity or the activation and deactivation of control policies. To capture this behavior, we employ an oscillatory transmission rate that generates recurrent epidemic waves of declining amplitude, providing a more challenging multi-wave setting in which to evaluate algorithmic performance.

\begin{description}
    \item[Example 1.] Synthetic data were generated over $T = 60$ time points with time-varying transmission rate $\beta_t = 0.3 \cdot \exp\big(\sin(2\pi t / 55) - t / 80\big)$. The incubation and infectious periods were set to $\alpha = 1/2$ and $\gamma = 1/7$, respectively. Each state ensemble was initialized using truncated normal distributions: $S_0\sim\mathcal{TN}_{[0,\, \infty]}(5\times10^5, 0.2^2) $,  $I_0\sim\mathcal{TN}_{[0,\, \infty]}(10, 0.2^2) $ and  $E_0=R_0=0$ .The priors were specified as $
\beta_0 \sim \mathcal{N}(0.3, 0.01^2),~
\alpha \sim \mathcal{TN}_{[0,\, \infty]}(0.6, 0.3^2), ~
\gamma \sim \mathcal{TN}_{[0,\, \infty]}(0.2, 0.1^2), ~
\nu_{\beta} \sim \mathcal{U}(0, 0.5).
$

 \item[Example 2.] Synthetic data were generated over $T=100$ time points with a time-varying transmission rate $
\beta_t = 0.5 \cdot \exp\Big(-(t - 15)^2/20^2\Big) + 0.065$,
and fixed parameters $\alpha = 1/3$ and $\gamma = 1/8$. The priors were specified as
$
\beta_0 \sim \mathcal{N}(0.35, 0.01^2), ~
\alpha \sim \mathcal{TN}_{[0,\, \infty]}(0.4, 0.2^2), ~
\gamma \sim \mathcal{TN}_{[0,\, \infty]}(0.12, 0.2^2), ~
\nu_{\beta} \sim \mathcal{U}(0, 0.3).
$
Initial state ensemble members were identical to Example 1.

\item[Example 3.] Synthetic data were generated over $T=100$ time points with an oscillatory, multi-wave transmission rate $\beta_t = 0.3 \cdot \exp\big(\cos(2\pi t / 70) - t/100\big)$, and fixed parameters $\alpha = 1/2$ and $\gamma = 1/6$. The priors were specified as
$
\beta_0 \sim \mathcal{N}(0.8, 0.01^2), ~
\alpha \sim \mathcal{TN}_{[0,\, \infty]}(0.6, 0.3^2), ~
\gamma \sim \mathcal{TN}_{[0,\, \infty]}(0.17, 0.05^2), ~
\nu_{\beta} \sim \mathcal{U}(0, 1).
$
A more informative prior was placed on $\gamma$ relative to Examples~1 and~2, as the oscillatory structure of $\beta_t$ introduces identifiability challenges that make the recovery rate harder to pin down from incidence data alone. Initial state ensemble members were identical to Examples~1 and~2.
\end{description}

For each simulation study, we compared the proposed eSMC$^2$ algorithm with the standard SMC$^2$ implementation that employs a bootstrap particle filter (BPF) in its inner layer (see Appendix \ref{appA}). In all cases, we used $N_\theta = 1000$ parameter particles and $N_x = 200$ ensemble members (or particles). For simplicity and comparability, the value of $N_x$ was kept fixed throughout the timeline. Parameter rejuvenation was triggered when the effective sample size of the parameter particles dropped below 50\% of $N_\theta$, followed by five iterations of the PMMH kernel.

\begin{figure}[H]
    \centering
    \includegraphics[width=1\linewidth]{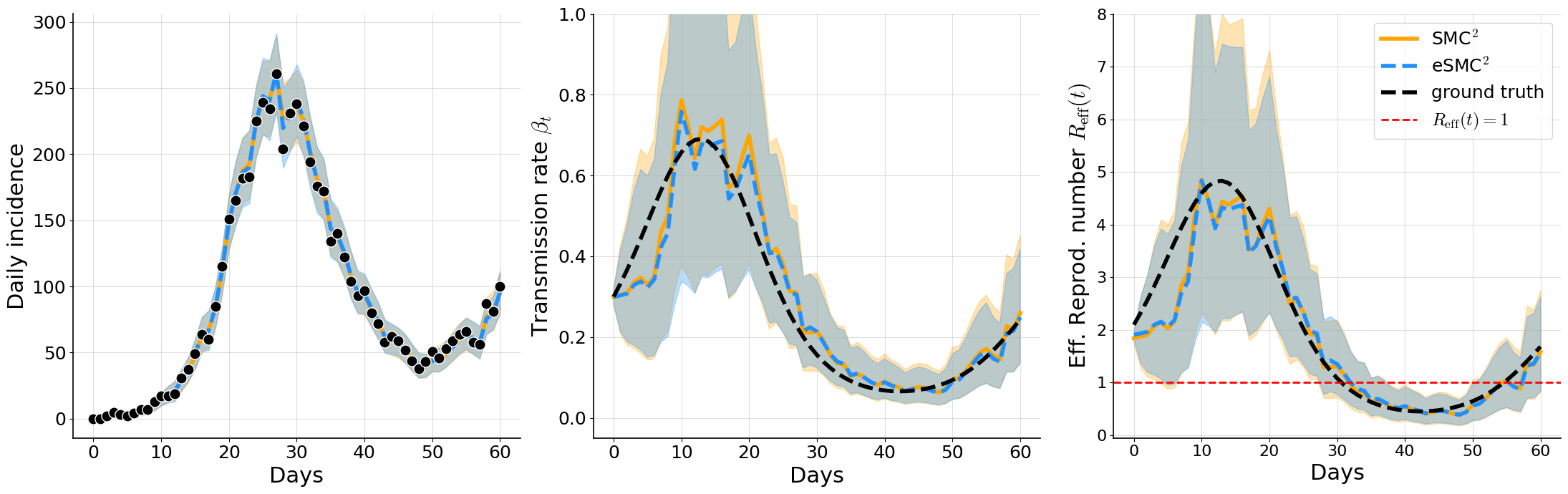}
    \caption{\footnotesize \textbf{Example 1: Filtered estimates of simulated incidence, transmission rate, and effective reproduction number.} 
    Solid orange lines (SMC$^2$) and dashed blue lines (eSMC$^2$) show posterior means; shaded areas indicate 95\% credible intervals. Observed incidence is shown as black dots, and the black dashed line represents the ground-truth latent quantity.}
    \label{Fig3}
\end{figure}

\begin{figure}[H]
    \centering
    \includegraphics[width=1\linewidth]{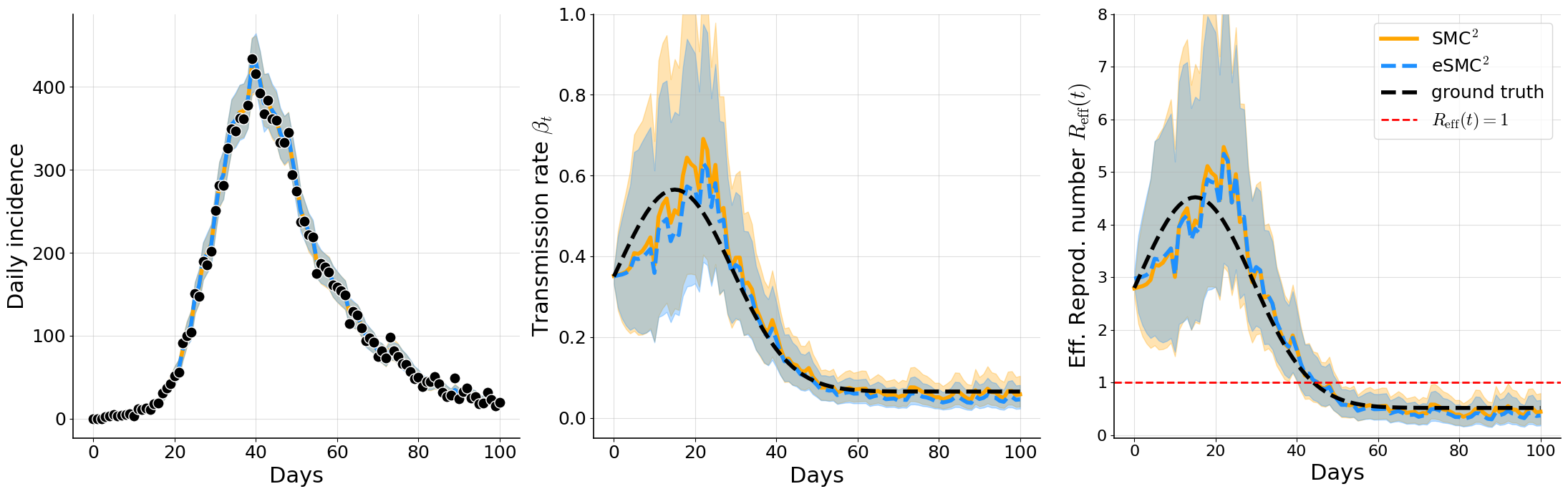}
    \caption{\footnotesize \textbf{Example 2: Filtered estimates of simulated incidence, transmission rate, and effective reproduction number.} 
    Solid orange lines (SMC$^2$) and dashed blue lines (eSMC$^2$) show posterior means; shaded areas indicate 95\% credible intervals. Observed incidence is shown as black dots, and the black dashed line represents the ground-truth latent quantity.}
    \label{Fig4}
\end{figure}

\begin{figure}[H]
    \centering
    \includegraphics[width=1\linewidth]{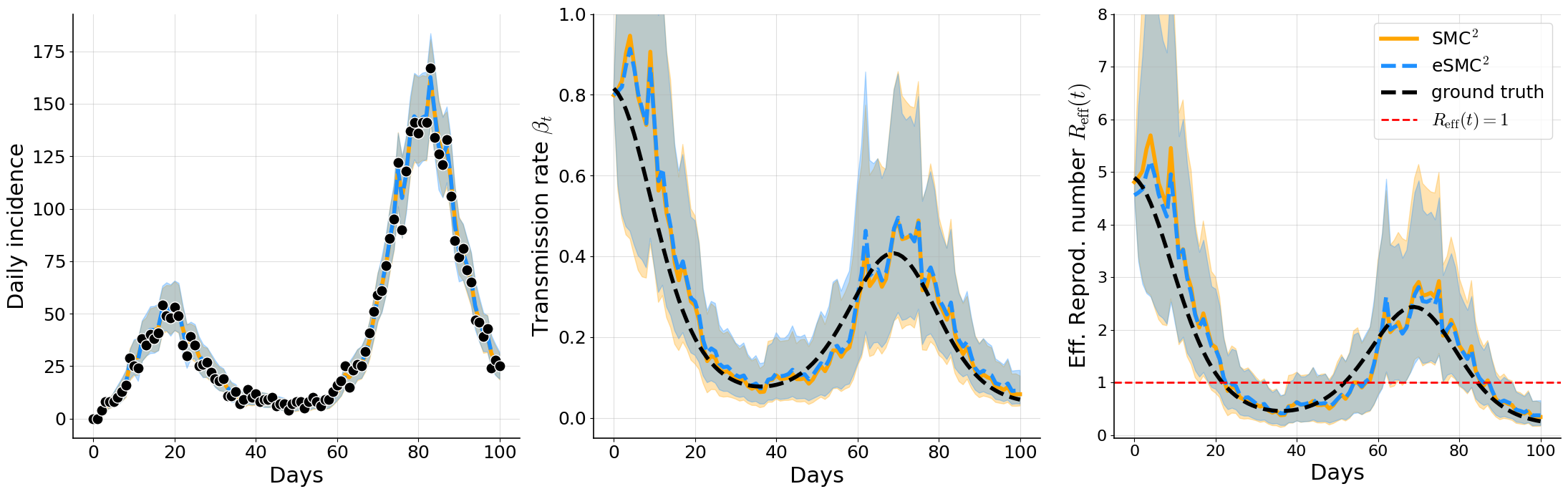}
    \caption{\footnotesize \textbf{Example 3: Filtered estimates of simulated incidence, transmission rate, and effective reproduction number.} 
    Solid orange lines (SMC$^2$) and dashed blue lines (eSMC$^2$) show posterior means; shaded areas indicate 95\% credible intervals. Observed incidence is shown as black dots, and the black dashed line represents the ground-truth latent quantity.}
    \label{Fig_ex3_state}
\end{figure}

Filtered estimates in Figs~\ref{Fig3}, ~\ref{Fig4}, and~\ref{Fig_ex3_state} correspond to a single representative run of each method. The results illustrate that both eSMC$^2$ (blue) and SMC$^2$ (orange) accurately capture the epidemic dynamics in Examples~1,~2, and~3. For both methods, the predicted incidence closely follows the observed data, with 95\% credible intervals consistently encompassing the actual case counts, indicating reliable uncertainty quantification. Estimates of the time-varying transmission rate $\beta_t$ and the effective reproduction number $R_{\text{eff}}(t)$ are broadly consistent across methods and closely align with the ground truth, demonstrating that both approaches yield comparable assessments of the evolving transmission dynamics. The wider credible intervals for $R_{\text{eff}}(t)$ relative to the filtered incidence reflect two complementary sources of uncertainty: the roughness of the random walk prior on $\beta_t$, and parameter non-identifiability, particularly during the early epidemic phase when different combinations of $\beta_t$, $\alpha$, and $\gamma$ can yield indistinguishable incidence curves. This wider uncertainty is therefore an accurate signal that incidence data alone cannot sharply identify instantaneous transmission intensity. Smoother priors on $\beta_t$, such as integrated Brownian motion or higher-order random walks, could potentially mitigate this variability but were not considered here and remain an avenue for future work. The filtering estimates of the unobserved latent state also closely track the true latent trajectories, with their credible intervals providing appropriate coverage (see Appendix \ref{appC}).

\begin{figure}[H]
    \centering
    \includegraphics[width=1\linewidth]{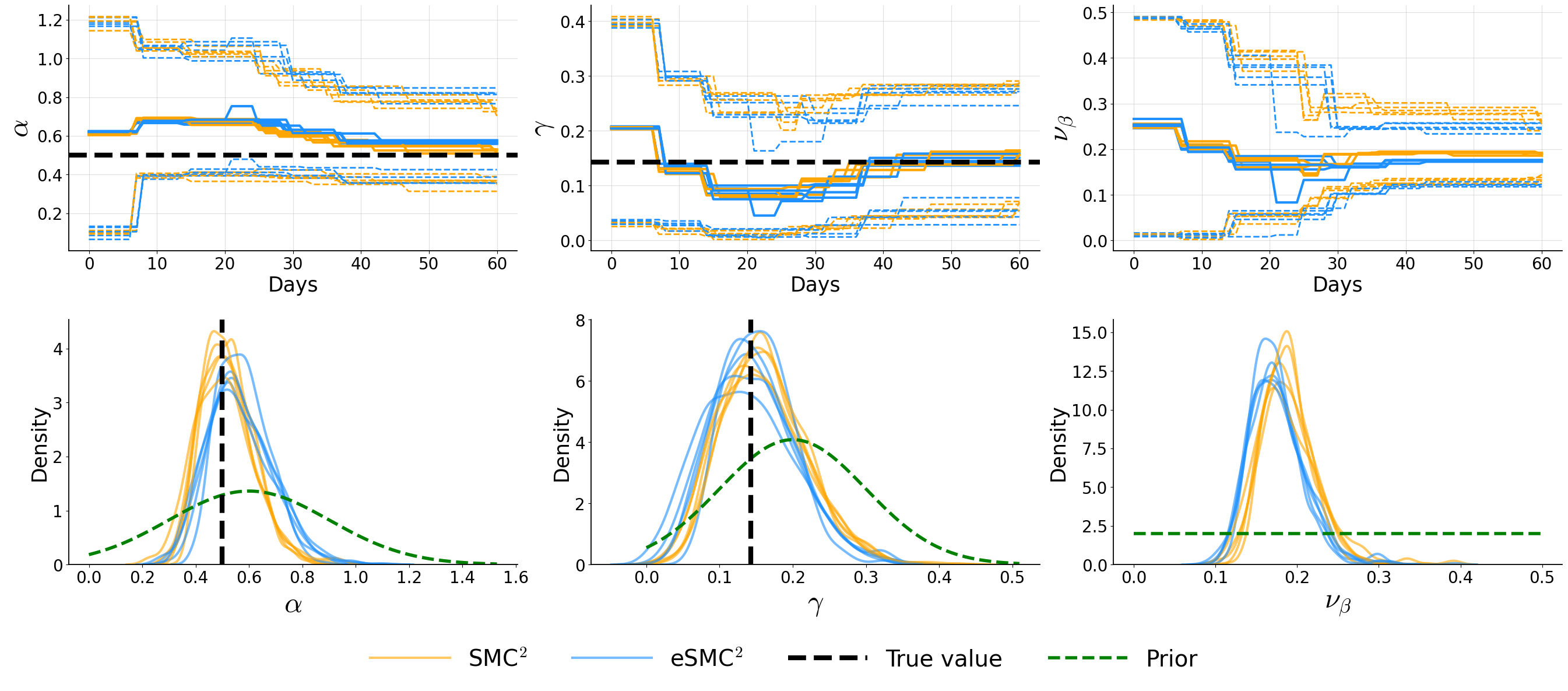}
  \caption{\footnotesize \textbf{Example 1: Posterior distributions of $\alpha$, $\gamma$, and $\nu_{\beta}$ for five independent runs.} Top row shows filtered means with 95\% credible intervals. 
Bottom row shows marginal posterior distributions at $T=60$, with prior distributions overlaid (green dashed lines). 
Black dashed lines indicate true parameter values.}

    \label{Fig5}
\end{figure}

\begin{figure}[H]
    \centering
    \includegraphics[width=1\linewidth]{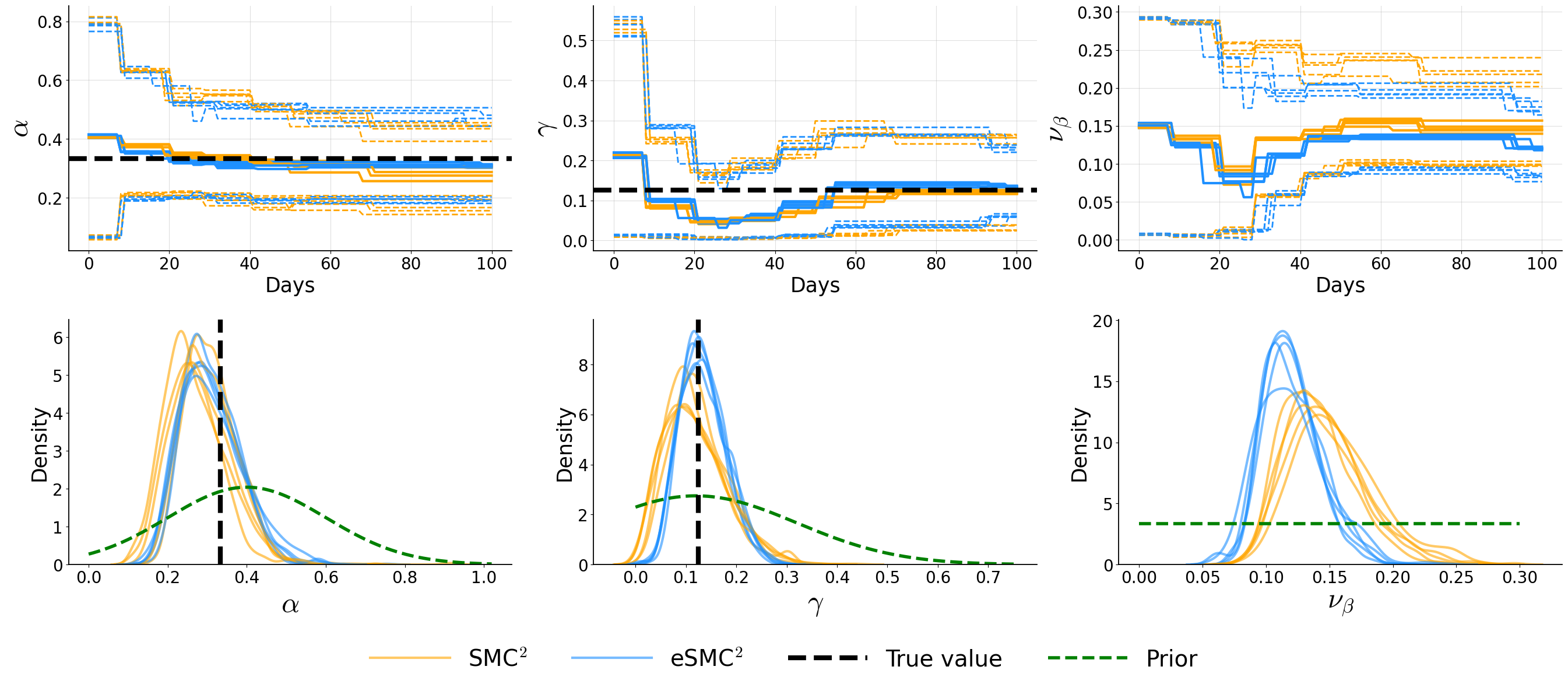}
    \caption{\footnotesize \textbf{Example 2: Posterior distributions of $\alpha$, $\gamma$, and $\nu_{\beta}$ for five independent runs.} The top row shows filtered means with 95\% credible intervals. 
Bottom row shows marginal posterior distributions at $T=100$, with prior distributions overlaid (green dashed lines). 
Black dashed lines indicate true parameter values.}
    \label{Fig6}
\end{figure}

\begin{figure}[H]
    \centering
    \includegraphics[width=1\linewidth]{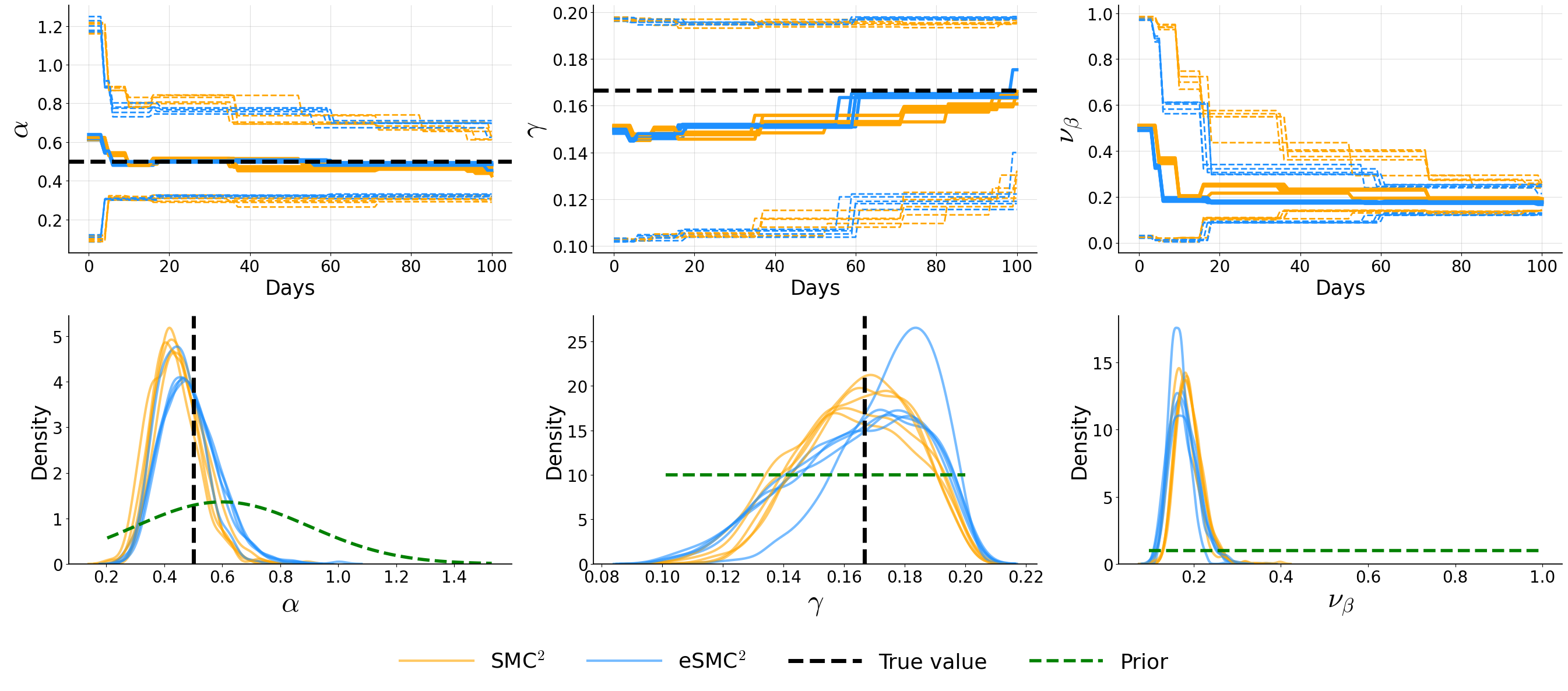}
    \caption{\footnotesize \textbf{Example 3: Posterior distributions of $\alpha$, $\gamma$, and $\nu_{\beta}$ for five independent runs.} The top row shows filtered means with 95\% credible intervals. 
Bottom row shows marginal posterior distributions at $T=100$, with prior distributions overlaid (green dashed lines). 
Black dashed lines indicate true parameter values.}
    \label{Fig_ex3_param}
\end{figure}

Figs~\ref{Fig5} and~\ref{Fig6} and~\ref{Fig_ex3_param} illustrate the temporal evolution of the posterior means of the inferred parameters, along with their marginal posterior densities at the final time step, each summarizing five independent runs. Across runs, both SMC$^2$ and eSMC$^2$ accurately recover the ground truth. For the latency and recovery rates ($\alpha$ and $\gamma$), the posterior means obtained from both algorithms are closely aligned, with credible intervals narrowing steadily as more data are assimilated, indicating progressive information gain. The only notable exception is the diffusion parameter $\nu_{\beta}$ in Examples 2 and 3, where a slightly larger discrepancy is observed. This difference likely arises from the stochastic nature of the log-transmission process combined with the approximation introduced by the EnKF-based likelihood in eSMC$^2$. Importantly, these differences do not introduce systematic bias or distort the inferred $\beta_t$ trajectory. The joint posterior contour plots (see Appendix \ref{appC}) further reveal strong agreement in posterior geometry and dependence structure: the contours for SMC$^2$ and eSMC$^2$ largely overlap, are centered near the true values, and display similar correlation patterns among parameters. The only visible difference appears along the $\nu_{\beta}$ dimension. Additional sensitivity analyses (Appendix \ref{appD}) assessing the effect of the number of state particles ($N_x$) and the reporting fraction ($\rho$) similarly demonstrate that posterior inferences are robust to ensemble size and exhibit only moderate variation when $\rho$ is correctly specified. The analyses also investigate the impact of non-informative (flat) priors, demonstrating that posterior estimates remain stable under weakly informative prior specifications.

Results in Table~\ref{Tab1} summarize the accuracy and computational cost of each method, averaged over five independent runs. The mean absolute error (MAE) and root mean squared error (RMSE) quantify discrepancies between the filtering means and the ground truth for both selected latent state ($Z_t,~\beta_t$)  and static parameters ($\alpha,~\gamma$). Across both examples, eSMC$^2$ achieves error magnitudes closely matching those of SMC$^2$, indicating that the ensemble approximation preserves inferential accuracy. In terms of computational effort, eSMC$^2$ offers a substantial advantage. In Example 1, the average CPU time is approximately 467 seconds for eSMC$^2$, compared with 2277 seconds for SMC$^2$, corresponding to a reduction by a factor of about  4.87. Similarly, in Example 2, eSMC$^2$ reduces average runtime to 639 seconds versus 3929 seconds for SMC$^2$, a factor of roughly 6.1. In Example~3, eSMC$^2$ completes in approximately 543 seconds on average, compared with 3766 seconds for SMC$^2$, corresponding to a speedup factor of roughly 6.9. The variability in the CPU across the run in both methods reflects the stochastic occurrence of rejuvenation steps, which introduce random fluctuations in computing time. 

\begin{table}[!ht]

\centering
\caption{\footnotesize\textbf{Comparison of goodness-of-fit and parameter estimates.} Each method is evaluated based on CPU time in seconds (with standard deviation) and MAE (RMSE in parentheses) for daily incidence, effective reproduction number, transmission rate and model static parameters. Results are shown for Examples 1, 2, and 3.}
\label{Tab1}

\small
\begin{tabular}{|l|c|c|c|c|c|c|}
\hline
\textbf{Method} & \textbf{CPU (sd)} & \textbf{Incidence ($\bm{\rho Z_t}$)}  & $\bm{R_{\text{eff}}(t)}$ & $\bm{\beta_t}$ & $\bm{\widehat{\mathrm{E}}[\alpha|y_{1:T}]}$ & $\bm{\widehat{\mathrm{E}}[\gamma|y_{1:T}]}$ \\
\hline
\multicolumn{7}{|l|}{\textbf{\textit{Example 1}}} \\ \hline
SMC$^2$   & 2277$\pm$145 & 3.257 (4.691) & 0.297 (0.455) & 0.044 (0.063) & 0.080 (0.101) & 0.047 (0.060) \\ \hline
eSMC$^2$  & 467$\pm$44   & 3.067 (4.156) & 0.309 (0.444) & 0.043 (0.062) & 0.103 (0.134) & 0.046 (0.057) \\ \hline

\multicolumn{7}{|l|}{\textbf{\textit{Example 2}}} \\ \hline
SMC$^2$  & 3929$\pm$433 & 4.652 (6.536) & 0.266 (0.450) & 0.032 (0.050) & 0.072 (0.087) & 0.049 (0.061) \\ \hline
eSMC$^2$ & 639$\pm$48   & 4.780 (6.697) & 0.233 (0.386) & 0.033 (0.053) & 0.064 (0.078) & 0.036 (0.046) \\ \hline

\multicolumn{7}{|l|}{\textbf{\textit{Example 3}}}\\ \hline
SMC$^2$ &3766$\pm$464 & 2.610 (3.752) & 0.308 (0.467) & 0.049 (0.073) & 0.055 (0.099) & 0.003 (0.019) \\ \hline
eSMC$^2$ & 543$\pm$33  & 2.444 (3.387) & 0.280 (0.393) & 0.047 (0.065) & 0.020 (0.098) & 0.004 (0.021)\\ \hline
\end{tabular}

\end{table}

As an additional benchmark, we evaluated eSMC$^2$ against the Liu and West filter \citep{liu2001combined} (using $10^4$ particles, runtime $\approx 9$ minutes) and a long PMCMC run ($10^4$ iterations, runtime $\approx 3$ hours) as a gold-standard reference, both serving as alternative approaches for joint sequential state-parameter estimation. Results provided in Appendix \ref{appE} demonstrate that eSMC$^2$ achieves superior parameter accuracy and produces narrower, more stable credible intervals compared to the Liu and West filter, while remaining in close agreement with the PMCMC posterior distributions evaluated at the last time point.

\subsection{Application to the 2022 monkeypox outbreak in the USA}\label{mpox}
We applied the proposed methodology to infer the transmission dynamics of the 2022 U.S. monkeypox (mpox) outbreak, caused by the West African clade II of the monkeypox virus. This outbreak represented an unprecedented global event, with the United States reporting among the highest case counts worldwide. The typical incubation period ranges from $3-21$ days, and infectiousness generally lasts $2-4$ weeks. The dataset consists of daily confirmed mpox case counts reported by the U.S. Centers for Disease Control and Prevention (CDC) from 6 May to 31 December 2022. Recent studies have applied Bayesian filtering to mpox data. For instance, \cite{saldana2023data} estimated the instantaneous reproduction number and growth rates using sequential Bayesian updating, while \cite{papageorgiou2024novel} proposed a particle filter incorporating penalty factors specific to mpox transmission. A related stochastic Kalman-based approach, validated on the 2022 Czech Republic mpox data, further discusses the challenges of Bayesian filtering for epidemic modeling with dynamic parameters \citep{papageorgiou2025new}. Our framework offers a complementary perspective by enabling scalable joint inference of latent epidemic states and model parameters, allowing uncertainty quantification directly from noisy daily surveillance data.

Given the pronounced day-to-day variability in the raw counts, we employed the Negative Binomial observation model as the primary specification. Preliminary analyses with an equidispersed (Poisson) observation model produced unstable posterior trajectories and parameter degeneracy, indicating that the Poisson assumption was inconsistent with the variability present in the daily data. The Poisson model becomes more plausible when incidence is aggregated over 7-day intervals, as temporal averaging reduces excess variability. Supporting results for the aggregated analysis are provided in Appendix \ref{appF}.

The algorithm was implemented with $N_x = 200$ ensemble members and $N_\theta = 1000$ parameter particles, values found to yield stable posterior estimates without excessive computational burden. All results and graphical outputs presented below were obtained using eSMC$^2$.

\begin{figure}[H]
    \centering
    \includegraphics[width=1\linewidth]{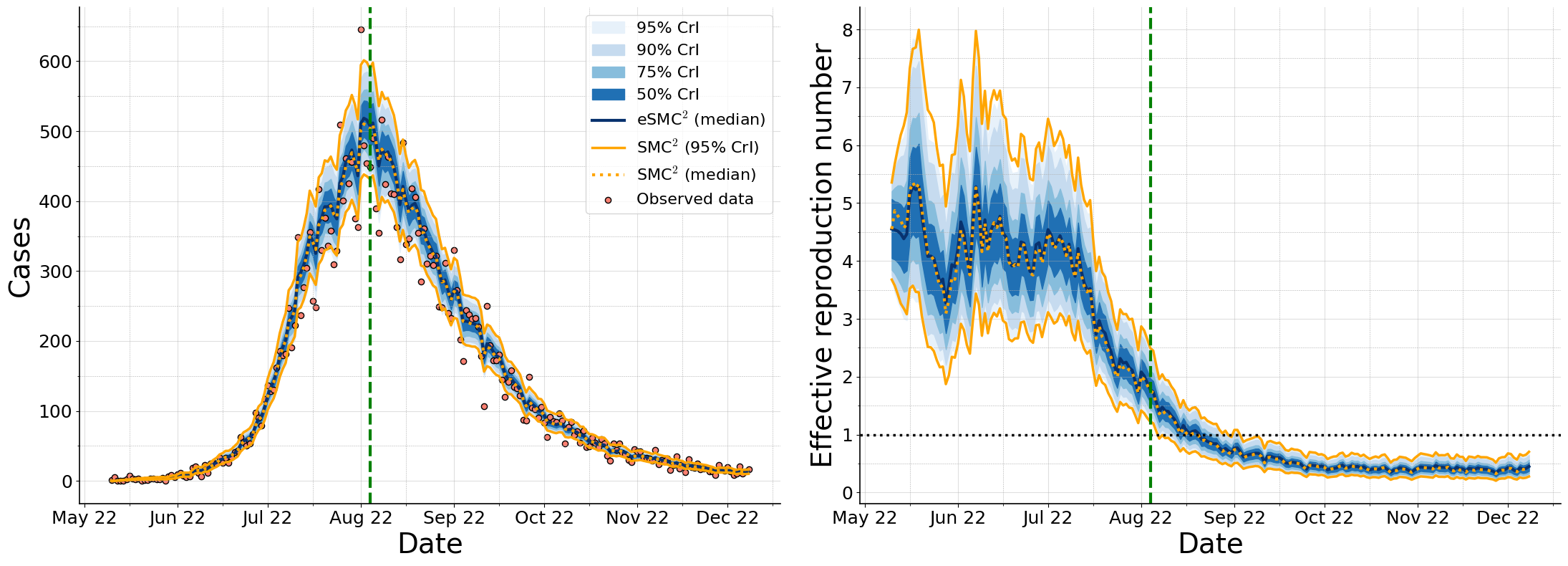}
    \caption{\footnotesize \textbf{Inference of daily incidence and effective reproduction number for the 2022 U.S. monkeypox outbreak.}
    Left: filtered estimates of daily incidence. The solid blue line denotes the posterior median obtained using eSMC$^2$, with reported case counts shown as red dots. 
    Right: inferred effective reproduction number $R_{\mathrm{eff}}(t)$.
    For eSMC$^2$, shaded regions represent 50\%, 75\%, 90\%, and 95\% credible intervals.
    The vertical dashed line indicates the declaration of the national state of emergency. Results from SMC$^2$ are shown in orange, with solid lines indicating the 95\% credible interval bounds and the dotted line representing the posterior median.}
    \label{Fig7}
\end{figure}

Fig.~\ref{Fig7} shows the estimated daily incidence and $R_{\mathrm{eff}}(t)$ over time. The filtered trajectories closely follow the observed case counts, while the credible bands reflect uncertainty arising from both stochastic epidemic dynamics and observation noise. The estimated $R_{\mathrm{eff}}(t)$ indicates sustained transmission above one during the early phase of the outbreak, followed by a decline after the declaration of the state of emergency, and eventual stabilisation below one as the outbreak subsides. For benchmarking purposes, we additionally report results obtained using the standard SMC$^2$ algorithm under identical model and prior specifications. In Fig.~\ref{Fig7}, SMC$^2$ results are shown using 95\% credible interval bounds (solid orange lines) together with the posterior median trajectory (dotted orange line), enabling a direct comparison of both central estimates and uncertainty quantification between the two methods.

\begin{figure}[H]
    \centering
    \includegraphics[width=1\linewidth]{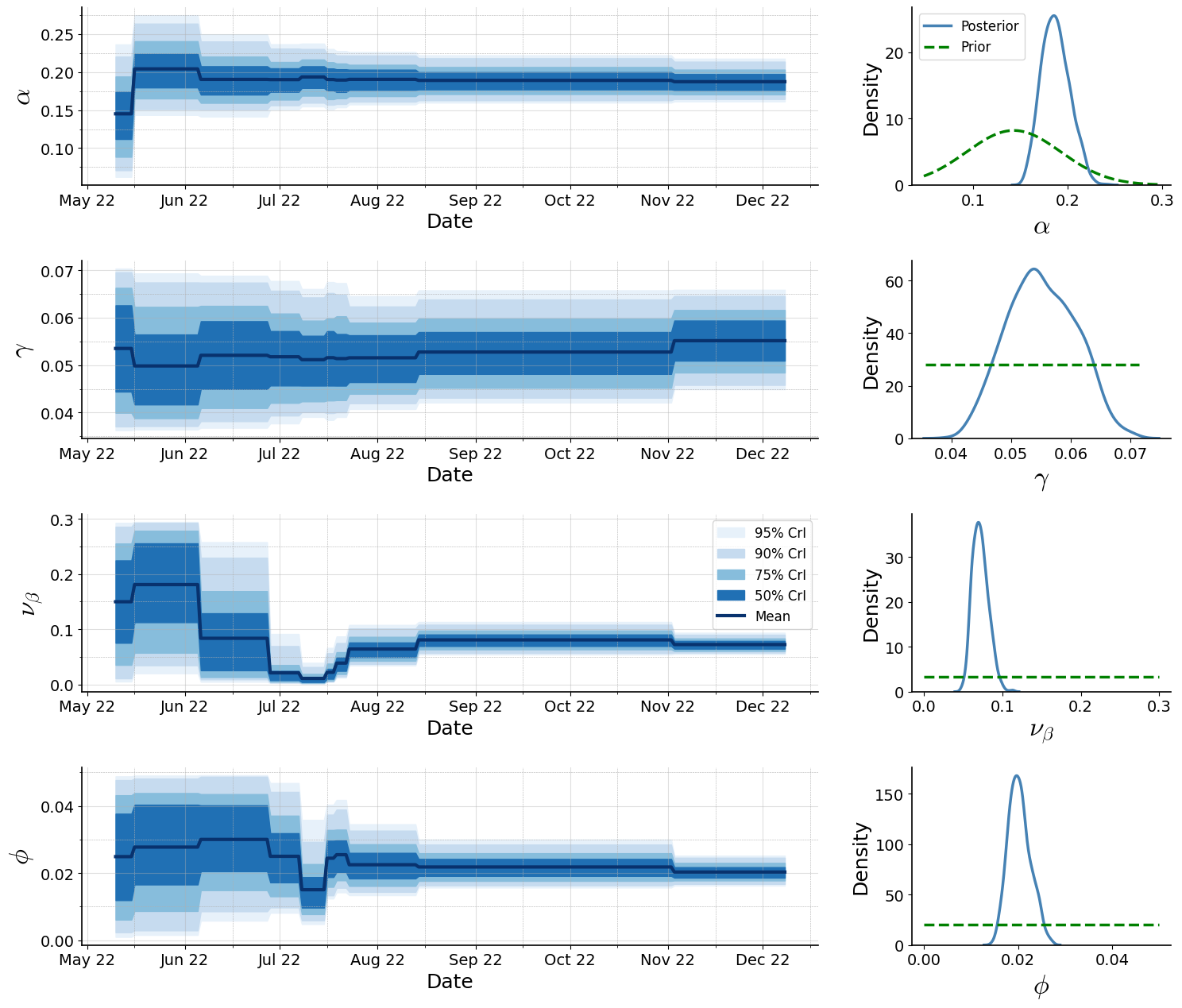}
\caption{\footnotesize \textbf{Posterior distributions of key epidemiological parameters.} The left column shows filtered means with credible intervals; the right column displays marginal posterior distributions at the last time step.}
\label{Fig8}
\end{figure}

Fig~\ref{Fig8} presents the posterior trajectories of key epidemiological parameters, with corresponding priors and posterior summaries reported in Table~\ref{Tab2}. As more data are assimilated, these parameters become increasingly well-identified, as reflected by the progressive narrowing of the posterior intervals. The analysis indicates a mean incubation period of approximately $5.3$ days (95\% CrI: $4.6$--$6.2$ days), consistent with early empirical estimates reported by \cite{madewell2023serial}, and a mean infectious period of approximately $18.2$ days (95\% CrI: $15.2$--$22.2$ days). The overdispersion parameter $\phi$ exhibits posterior mass distinctly separated from zero, confirming substantial extra-Poisson variability. For comparison, we also implemented the standard SMC$^2$ algorithm under identical model and prior settings. Results reported in Appendix \ref{appF} show that both methods produce a very similar evolution of the filtering distributions for the parameters, indicating that the ensemble-based approximation in eSMC$^2$ accurately captures posterior uncertainty. Notably, a single run with these settings required approximately 16 minutes of CPU time, compared with nearly 2.5 hours for the equivalent SMC$^2$ implementation, demonstrating the efficiency gains achieved by the EnKF-based approach.

\begin{table}[!ht]
\centering
\caption{\footnotesize\textbf{Description of the different parameters, priors, and posterior estimates at the final time step.} Upper and/or lower bounds have been imposed by the observations. $\mathcal{TN}$ denotes a truncated normal distribution $\mathcal{TN}_{[\text{inf, sup}]}(\text{mean}, \text{std}^2)$.}
\label{Tab2}

\small
\begin{tabular}{|c|l|c|c|}
\hline
\textbf{Parameter} & \textbf{Description} & \textbf{Prior/Value} & \textbf{Posterior mean (95\% CrI)} \\ \hline

$\beta_{0}$ & Initial condition transmission rate 
& $\mathcal{U}(0.2,\,0.3)$ 
& -- \\ \hline

$\alpha$ & Latency rate ($1/\alpha$ incubation period) 
& $\mathcal{TN}_{[1/21,\,1/3]}(1/7,\,0.05^2)$ 
& 0.187 (0.161, 0.218) \\ \hline

$\gamma$ & Recovery rate ($1/\gamma$ infectious period) 
& $\mathcal{U}(1/28,\,1/14)$ 
& 0.055 (0.045, 0.066) \\ \hline

$\nu_{\beta}$ & Volatility of the Brownian process
& $\mathcal{U}(0,\,0.3)$ 
& 0.072 (0.055, 0.094) \\ \hline

$\phi$ & Overdispersion 
& $\mathcal{U}(0,\,0.05)$ 
& 0.020 (0.016, 0.025) \\ \hline

\end{tabular}

\end{table}

\begin{figure}[H]
    \centering
    \includegraphics[width=1\linewidth]{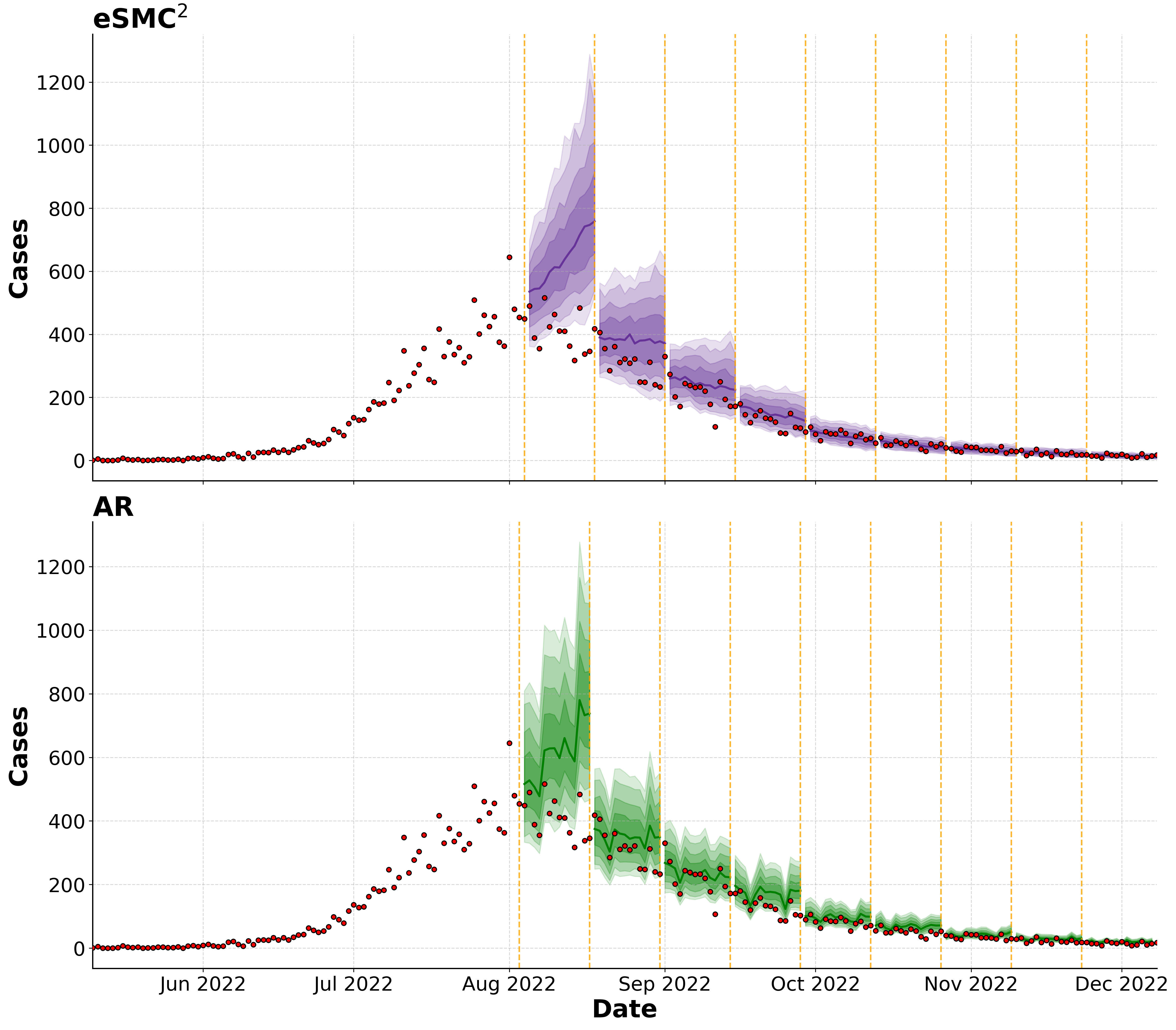}
\caption{\footnotesize \textbf{Posterior-predictive forecasts at different starting dates.} The top panel shows eSMC$^2$ forecasts, and the bottom panel shows AR forecasts. Median forecasts are displayed in purple (eSMC$^2$) and green (AR), with corresponding 50\%, 75\%, 90\%, and 95\% predictive intervals. Observed case counts are shown in red. Vertical orange dashed lines indicate the forecast start dates.}
\label{Fig9}
\end{figure}

Fig~\ref{Fig9} presents 14-day-ahead posterior-predictive forecasts initiated at multiple time points during the outbreak, assuming the transmission rate $\beta_t$ remains fixed at its last estimated value. Forecasts are obtained by propagating the latent state forward through the state-space dynamics and sampling synthetic observations from the observation model. The purpose of this analysis is to evaluate the short-term predictive performance of the proposed eSMC$^2$ framework. We compare the results with an autoregressive (AR) model (Appendix~F). The top panel of Fig.~\ref{Fig9} shows forecasts from eSMC$^2$ and the bottom panel those from the AR model. Quantitative performance is summarised in Table~\ref{tab:forecast_eval}, which reports mean absolute error (MAE), weighted interval score (WIS; based on the 50\%, 75\%, 90\%, and 95\% intervals), and empirical 95\% prediction interval (PI) coverage for each forecast start date. Both models struggle to anticipate the turning point in early August, and around this period the autoregressive model achieves a lower forecast error. In particular, around the epidemic peak, the AR model consistently outperforms eSMC$^2$ across all evaluation metrics. This likely reflects the intrinsic difficulty of forecasting near the peak, where rapid changes in transmission dynamics and the assumption of a fixed $\beta_t$ limit the responsiveness of the mechanistic model. In contrast, the AR model, being purely data-driven, can more readily adapt to short-term trends during this highly dynamic phase. Similar behaviour has been reported in related work, where data-driven Hawkes-type processes were found to outperform stochastic SEIR models during abrupt regime shifts \citep{temfack2025sequential}. However, from mid-September onward, eSMC$^2$ achieves lower MAE and WIS in most forecast windows and maintains coverage close to the nominal 95\% level, whereas the autoregressive model exhibits declining coverage in later periods (Table~\ref{tab:forecast_eval}). These results highlight a limitation of the current modelling setup in capturing abrupt turning points, mainly due to the inertia of the mechanistic dynamics, rather than an inherent limitation of SMC-based inference. At the same time, they confirm the strength of the approach in providing well-calibrated probabilistic forecasts once the epidemic enters a more stable phase. Long-term forecasts for the final epidemic phase are presented in Fig.~\ref{Fig10}. The model reproduces the overall decline and achieves full 95\% predicted interval coverage.

\begin{table}[!ht]
\centering
\caption{\footnotesize\textbf{Evaluation of the 2 weeks-ahead forecasts of reported mpox cases during the 2022 U.S. outbreak}. Bold indicates better-performing model.}
\label{tab:forecast_eval}
\small
\begin{tabular}{|l|r|r|r||r|r|r|}
\hline
 & \multicolumn{3}{c||}{\textbf{eSMC$^2$}} & \multicolumn{3}{c|}{\textbf{AR}} \\
\cline{2-7}
\textbf{Forecast date} & \textbf{95\% PI coverage} & \textbf{MAE} & \textbf{WIS} & \textbf{95\% PI coverage} & \textbf{MAE} & \textbf{WIS} \\
\hline
2022-08-04 & 0.43 & 230.6 & 269.5 & \textbf{0.57} & \textbf{204.2} & \textbf{251.6} \\
\hline
2022-08-18 & \textbf{1.00} & 78.5 & 121.3 & \textbf{1.00} & \textbf{52.1} & \textbf{94.2} \\
\hline
2022-09-01 & 0.86 & 41.2 & 74.5 & \textbf{0.93} & \textbf{30.8} & \textbf{60.3} \\
\hline
2022-09-15 & \textbf{1.00} & \textbf{25.4} & \textbf{43.9} & 0.79 & 41.2 & 55.5 \\
\hline
2022-09-29 & \textbf{1.00} & \textbf{10.7} & \textbf{22.6} & 0.86 & 15.2 & 25.2 \\
\hline
2022-10-13 & \textbf{1.00} & \textbf{7.8} & \textbf{15.2} & 0.79 & 18.3 & 22.8 \\
\hline
2022-10-27 & \textbf{1.00} & \textbf{5.2} & \textbf{11.2} & 0.86 & 7.5 & 11.7 \\
\hline
2022-11-10 & \textbf{1.00} & \textbf{4.6} & \textbf{8.5} & 0.64 & 7.3 & 9.6 \\
\hline
2022-11-24 & \textbf{0.93} & \textbf{3.9} & \textbf{6.5} & 0.57 & 5.9 & 7.7 \\
\hline
\end{tabular}
\end{table}

\section{Discussion and limitations}\label{sec: discussion}
In this work, we investigated sequential Bayesian inference for latent states and parameters in compartmental models observed through incidence data. We proposed the eSMC$^2$, an extension of the SMC$^2$ framework that replaces the particle filter with an EnKF-derived likelihood, enabling computationally efficient sequential updates of both latent epidemic trajectories and model parameters.  This approach builds on the eMCMC method of \cite{drovandi2022ensemble} but extends it to a fully sequential setting, allowing time-resolved inference in dynamic epidemic systems. By leveraging the ensemble approximation of the likelihood, eSMC$^2$ reduces the computational burden while retaining the ability to capture nonlinear state evolution. Within the broader landscape of Bayesian calibration approaches, including likelihood-free methods such as ABC \citep{li2025advances}, eSMC$^2$ represents a complementary strategy. Likelihood-free approaches are particularly useful when the likelihood is intractable or prohibitively expensive to evaluate. In contrast, eSMC$^2$ relies on an explicit state-space formulation and an approximated likelihood, enabling principled sequential updating of both latent states and parameters as new data arrive. Consequently, eSMC$^2$ is especially suited to time-series epidemic models where inference is performed sequentially, whereas likelihood-free methods may be preferable for models that do not admit a tractable state-space representation.

Methodologically, the EnKF provides a scalable alternative to standard particle filtering, particularly in moderately high-dimensional or computationally constrained settings. While the EnKF likelihood is only a Gaussian approximation, we employ an unbiased correction, following  \cite{drovandi2022ensemble}, which reduces sensitivity to the ensemble size. In addition, rather than assuming a fixed observation noise variance, we dynamically update the observation noise variance at each filtering step using the ensemble-predicted state estimates. This adaptive treatment of observation uncertainty allows the filter to better reflect time-varying variability in incidence data. Although we do not provide formal bias quantification, previous studies indicate that EnKF-based likelihoods yield stable and accurate inference in moderately nonlinear systems when combined with MCMC \citep{khalil2015estimation, drovandi2022ensemble}. Our simulation studies confirm this behavior, showing that the approximation error remains small and posterior means closely match those obtained with standard SMC$^2$.  Importantly, the diffusion-driven SEIR formulation combined with Bayesian filtering provides a flexible mechanism for detecting temporal changes in transmission, such as those resulting from behavioral shifts or intervention measures, without requiring explicit change-point modeling. In our simulations, the method accurately recovers both the transmission rate and the effective reproduction number, even without knowledge of their true trajectories. The application to the 2022 U.S. mpox outbreak further demonstrates that eSMC$^2$ delivers well-calibrated, near-real-time estimates of transmission dynamics and effective reproduction numbers, supporting timely situational awareness and public health decision-making. It should be noted that this orecasting analysis is retrospective and does not account for real-time complexities such as data revisions and reporting delays, so predictive performance may be optimistic relative to an operational setting \citep{charniga2024nowcasting}.

Despite its practical and computational advantages, several methodological and theoretical limitations should be noted. Unlike standard SMC$^2$, which is theoretically exact in the limit of infinite state and parameter particles, eSMC$^2$ inherits bias from the EnKF likelihood approximation. This bias may become more pronounced in strongly nonlinear or multimodal epidemic systems, and rigorous convergence guarantees for nonlinear, non-Gaussian settings remain an open challenge. While convergence of the EnKF likelihood has been established in linear-Gaussian systems with rate $\mathcal{O}(N_x^{-1/2})$ \citep{chen2022autodifferentiable}, extending such results to nonlinear systems is nontrivial.  Although the proposed method is sequential, it is not fully ``online'', as each call of the PMMH kernel still involves running the EnKF on all data available up to that point. One potential avenue to further improve computational efficiency would be to develop a fully recursive implementation analogous to a Nested Particle Filter (NPF) \citep{crisan2018nested}, in which the inner particle filter could be replaced by an EnKF. From a modeling standpoint, our SEIR application relies on a single data stream (daily incidence). Incorporating additional observations such as hospitalizations, deaths, or serological survey data could improve parameter identifiability, particularly in disentangling reporting rates from transmission parameters \citep{swallow2022challenges}. More generally, epidemiological models often include parameters constrained to the unit interval (e.g., reporting or mortality ratios). In such cases, one may either assign a \textit{Beta} prior, which naturally enforces support on $(0,1)$, or apply a \textit{logit} transformation and conduct inference on the unconstrained scale. Finally, while the framework is broadly applicable to other state-space models beyond epidemiology, its performance may vary depending on the degree of nonlinearity, the suitability of the Gaussian approximation for the forecast ensemble and the choice of observation noise. Future research will explore these limitations in more detail.

\section*{Acknowledgments}
This publication has emanated from research conducted with the financial support of Taighde
Éireann – Research Ireland under Grant number 21/FFP-P/10123. 

\section*{Data availability}
All code used to produce the analyses in this paper is available at  
\href{https://github.com/Dhorasso/bma-smc2-dthp-seir}{https://github.com/Dhorasso/bma-smc2-dthp-seir}.  
The datasets used are publicly available from the HPSC at  
\href{https://respiratorydisease-hpscireland.hub.arcgis.com/pages/influenza}{https://respiratorydisease-hpscireland.hub.arcgis.com/pages/influenza}  
for influenza data, and from the Irish COVID-19 Data Hub at  
\href{https://COVID19ireland-geohive.hub.arcgis.com/}{https://COVID19ireland-geohive.hub.arcgis.com/}  
for COVID-19 data.

\appendix
\renewcommand{\thefigure}{\thesection\arabic{figure}}
\renewcommand{\thetable}{\thesection\arabic{table}}
\renewcommand{\thealgorithm}{\thesection\arabic{algorithm}}
\counterwithin{figure}{section}
\counterwithin{table}{section}

\section{Transition density and observation error variance}\label{appA}

\subsection{Transition density of the SEIR model}\label{appA:sec1}
We discretize the continuous-time SEIR dynamics using a forward Euler scheme for the compartmental states $(S_t, E_t, I_t, R_t, Z_t)$ and an Euler–Maruyama scheme for the stochastic transmission rate $\log (\beta_t)$. The compartments are propagated deterministically according to
\begin{align}\label{disc_seir}
\begin{cases}
S_t = S_{t-1} - \Delta t \, \beta_{t-1} \dfrac{S_{t-1} I_{t-1}}{N},\\[1mm]
E_t = E_{t-1} + \Delta t \, \Big(\beta_{t-1} \dfrac{S_{t-1} I_{t-1}}{N} - \alpha E_{t-1} \Big),\\[1mm]
I_t = I_{t-1} + \Delta t \, (\alpha E_{t-1} - \gamma I_{t-1}),\\[1mm]
R_t = R_{t-1} + \Delta t \, \gamma I_{t-1},\\[1mm]
Z_t = \alpha E_{t-1} \Delta t.
\end{cases}
\end{align}
where $\Delta t$ represents the discretization step (one day in our implementation). Formally, this deterministic update can be represented as a Dirac delta distribution in the transition density:
\begin{align}
p(S_t, E_t, I_t, R_t, Z_t| x_{t-1}, \theta) = \delta_{f(x_{t-1}, \theta)}(S_t, E_t, I_t, R_t, Z_t),
\end{align}
where $f(\cdot)$ is the Euler update map in \eqref{disc_seir}.  The stochastic log-transmission rate is propagated as
\begin{align}
\log (\beta_t) = \log (\beta_{t-1}) + \nu_\beta \sqrt{\Delta t} \, \epsilon_t, \quad \epsilon_t \sim \mathcal{N}(0,1),
\end{align}
so that the full transition density for the latent state $x_t$ is given by
\begin{align}
p(x_t | x_{t-1}, \theta) = \delta_{f(x_{t-1}, \theta)}(S_t, E_t, I_t, R_t, Z_t)\, 
\mathcal{N}(\log (\beta_t); ~\log (\beta_{t-1}), ~\nu_\beta^2).
\end{align}

\subsection{Kalman gain and observation error variance} \label{appA:sec2}

We consider the stochastic process $(x_t, y_t)_{t \ge 0}$ defined in Section 2 of the main text, such that $y_t | x_t \sim p(y_t | x_t)$ with finite conditional mean $\mathrm{E}[y_t | x_t] = H x_t$. Then
\begin{align}\label{uncorrelated_residual}
 \mathrm{E}\Big[ \big(x_t - \mathrm{E}[x_t]\big) \big(y_t - \mathrm{E}[y_t | x_t]\big) \Big] 
 &= \mathrm{E} \Big[ \mathrm{E}\big[ \big(x_t - \mathrm{E}[x_t]\big) \big(y_t - \mathrm{E}[y_t | x_t]\big) \,\big|\, x_t \big] \Big] \notag\\ 
 &= \mathrm{E} \Big[ \big(x_t - \mathrm{E}[x_t]\big) \underbrace{\mathrm{E}\big[ y_t - \mathrm{E}[y_t | x_t] \,\big|\, x_t \big]}_{=0} \Big] = 0.
\end{align}

Equation~\eqref{uncorrelated_residual} implies that the deviations of the latent state are uncorrelated with the observation residuals, where the residuals are defined relative to their conditional mean. This result generalizes Lemma 1 of \citet{ebeigbe2020poisson} to any observation distribution with a finite conditional mean. We stress that this property does not imply that a linear Kalman gain is optimal in a nonlinear setting. Nevertheless, it justifies using a linear gain $K_t$ to update the state estimate:
\begin{align} 
K_t &= \mathrm{Cov}\big[x_t, y_t\big] \big(\mathrm{Var}[y_t]\big)^{-1} \notag\\
&= \mathrm{Cov}\big[x_t, \mathrm{E}[y_t | x_t] + y_t- \mathrm{E}[y_t | x_t]\big] \big(\mathrm{Var}[y_t]\big)^{-1} \notag\\
 &= \mathrm{Cov}\big[x_t, \mathrm{E}[y_t | x_t]\big] \big(\mathrm{Var}[y_t]\big)^{-1} \notag\\
 &= \mathrm{Cov}\big[x_t, \mathrm{E}[y_t | x_t]\big] \Big(\mathrm{Var}\big[\mathrm{E}[y_t | x_t]\big] + \mathrm{E}\big[\mathrm{Var}[y_t | x_t]\big]\Big)^{-1} \notag\\
 &\approx \widehat{\Sigma}_{t|t-1} H^\top \left( H \widehat{\Sigma}_{t|t-1} H^\top +  \underbrace{\frac{1}{N_x} \sum_{i=1}^{N_x} \mathrm{Var}\big[y_t | x_t^{(f,i)}}_{V_t^{N_x}}\big] \right)^{-1}.
\end{align}

The third line follows because the residual $(y_t - \mathrm{E}[y_t | x_t])$ is uncorrelated with $x_t$, as shown in \eqref{uncorrelated_residual}. The fourth line is a direct application of the law of total variance and the last line is an ensemble approximation of the Kalman gain, where $\widehat{\Sigma}_{t|t-1}$ is the forecast ensemble covariance.


\section{Bootstrap Particle Filter} \label{appB}

An unbiased estimate of the incremental likelihood can be obtained using a particle filter. Algorithm~\ref{alg:bpf} outlines the Bootstrap Particle Filter (BPF), which is the version employed in this paper within the standard SMC$^2$ framework.  Algorithm~\ref{alg:stratified-resampling} describes the stratified resampling procedure, which is used in both SMC$^2$ and eSMC$^2$ implementations.
\begin{algorithm}[H]
\caption{Bootstrap Particle Filter (BPF)} \label{alg:bpf}
Operations involving index $i$ must be performed for $i = 1,\dots, N_x$.

The indices $a_{t}^{1:N_x}$ define the ancestral state particles at time $t$ after the resampling.

\textbf{Inputs:} Observation: $y_{1:T}$, Number of particles: $N_x$, Inital state distribution: $p(x_{0})$, Parameter vector $\theta$.

\textbf{Output:} Particles set: $\left\{x^{i}_{0:t}, w^{i}_{ 0:t}\right\}_{i=1}^{N_x}$, marginal likelihood  $\widehat{ p}_{\text{bpf}}^{N_x}(y_{1: t}|\theta)$

\hrulefill 
\begin{algorithmic}[1]
    \State Sample initial  particles : $x^{i}_{0} \sim p( x_{0})$ 
    \State Compute weights:  $w^{i}_0 = 1$, \quad $W^{i}_{0} = 1/N_x$ 
    \For{$t=1$ to $T$}
        \State\label{res1} Sample new indices: $a_{t}^{1:N_x}\sim\text{Resample}(W_{t-1}^{1: N_x})$ \Comment{Algorithm~\ref{alg:stratified-resampling}}
        \State\label{fk} Propagate states $x_{t}^{i} \sim p(\cdot |x_{0:t-1}^{a_{t}^{i}},\theta)$
        \State\label{wpmf} Compute  weights and normalize:  $w_{t}^{i}=p(y_t |x_{t}^{i},\theta)$
        , \quad $W_{t}^{i} = w_{t}^{i}/\sum_{j=1}^{N_x}w^{j}_{t}$
        \State\label{inc} Compute the incremental likelihood:  $\widehat{ p}_{\text{bpf}}^{N_x}(y_t|y_{1: t-1},\theta)=\frac{1}{N_x}\sum_{i=1}^{N_x} w^{i}_{t}$
    \EndFor
\State Compute marginal likelihood: $   \widehat{p}_{\text{bpf}}^{N_x}(y_{1:T} |, \theta) = \prod_{t=1}^{T} \widehat{p}_{\text{bpf}}^{N_x}(y_t | y_{1:t-1}, \theta).$
\end{algorithmic}
\end{algorithm}

\begin{algorithm}[H]
\caption{Stratified Resampling} 
\label{alg:stratified-resampling}

\textbf{Inputs:} Normalized weights $W^{1:N_x}$, number of particles $N_x$.  

\textbf{Output:} Resampled indices $a^{1:N_x}$.

\hrulefill 

\begin{algorithmic}[1]
\State Compute cumulative weights: $C_i = \sum_{j=1}^i W^j$ for $i = 1,\dots,N_x$
\For{$k = 1,\dots,N_x$}
    \State Sample $u_k \sim \text{Uniform}\!\left(\tfrac{k-1}{N_x}, \tfrac{k}{N_x}\right)$
\EndFor
\State Set $i \gets 1$
\For{$k = 1,\dots,N_x$}
    \While{$u_k > C_i$}
        \State $i \gets i+1$
    \EndWhile
    \State $a^k \gets i$
\EndFor
\end{algorithmic}
\end{algorithm}

\section{Additional results from simulated experiments}\label{appC}

\begin{figure}[H]
    \centering
    \includegraphics[width=0.85\linewidth]{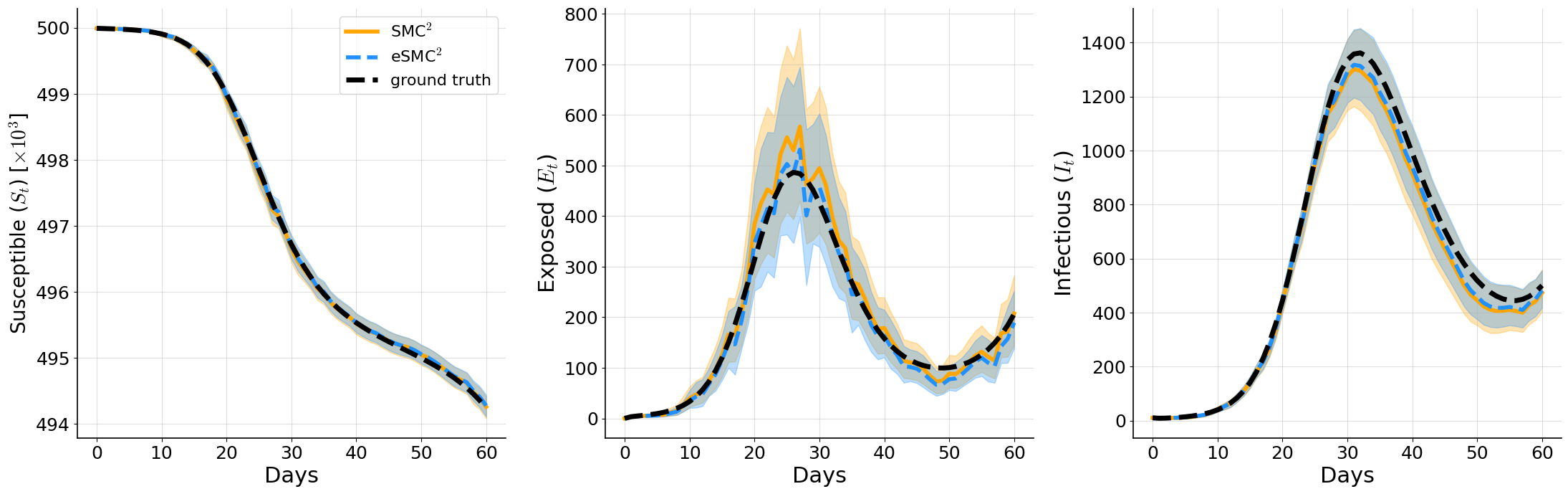}
   \caption{\footnotesize \textbf{Example 1: Unobserved states of the SEIR model.} Filtering mean and 95\% credible interval are shown for each compartments. The black dashed line indicates the ground truth.}
    \label{fig:unobs1}
\end{figure}

\begin{figure}[H]
    \centering
    \includegraphics[width=0.85\linewidth]{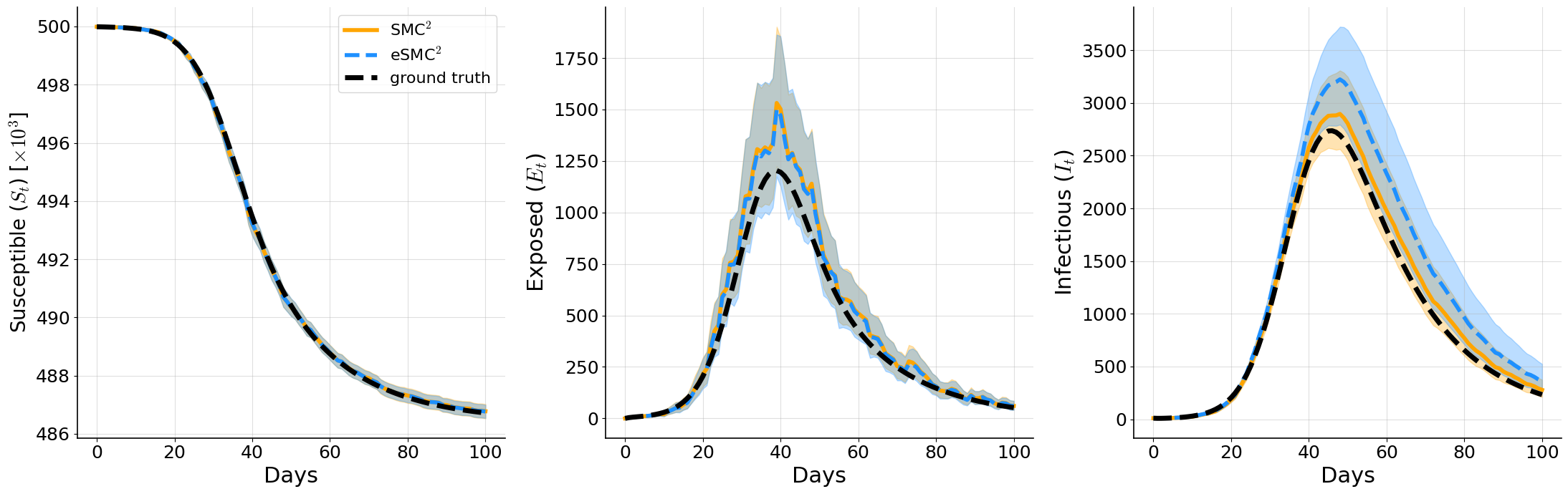}
   \caption{\footnotesize \textbf{Example 2: Unobserved states of the SEIR model.} Filtering mean and 95\% credible interval are shown for each compartments. The black dashed line indicates the ground truth.}
    \label{fig:unobs2}
\end{figure}

\begin{figure}[H]
    \centering
    \includegraphics[width=0.85\linewidth]{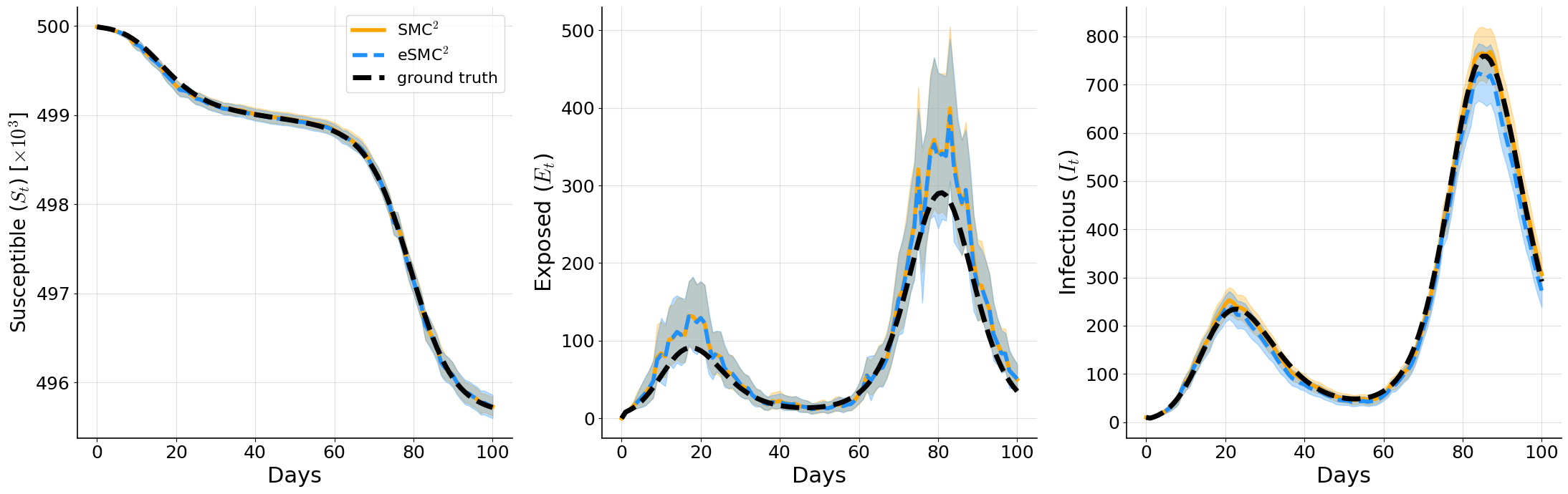}
   \caption{\footnotesize \textbf{Example 3: Unobserved states of the SEIR model.} Filtering mean and 95\% credible interval are shown for each compartments. The black dashed line indicates the ground truth.}
    \label{fig:unobs3}
\end{figure}

\begin{figure}[H]
    \centering
    \includegraphics[width=0.8\linewidth]{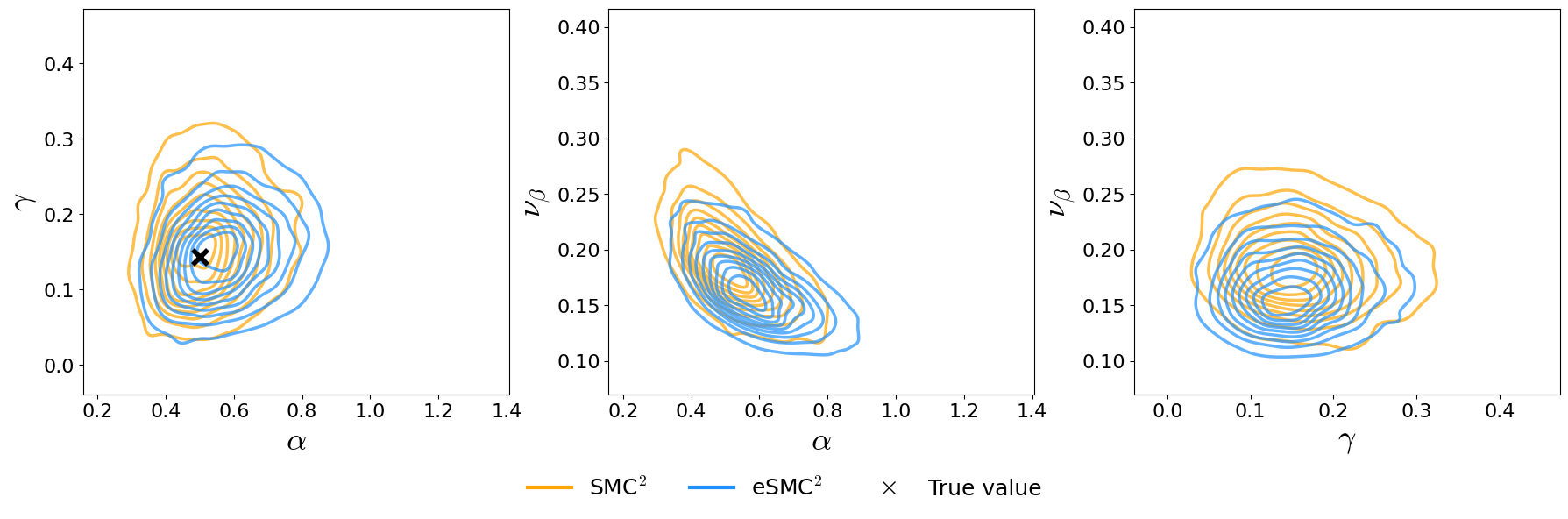}
\caption{\footnotesize \textbf{Posterior distributions of model parameters for Example 1.} Each subplot shows the pairwise marginal densities of $(\alpha, \gamma)$, $(\alpha, \nu_{\beta})$, and $(\gamma, \nu_{\beta})$. Contours represent the combined posterior from five runs of SMC$^2$ (orange) and eSMC$^2$ (blue), with black crosses indicating the true parameter values.}

    \label{fig:contour1}
\end{figure}

\begin{figure}[H]
    \centering
    \includegraphics[width=0.8\linewidth]{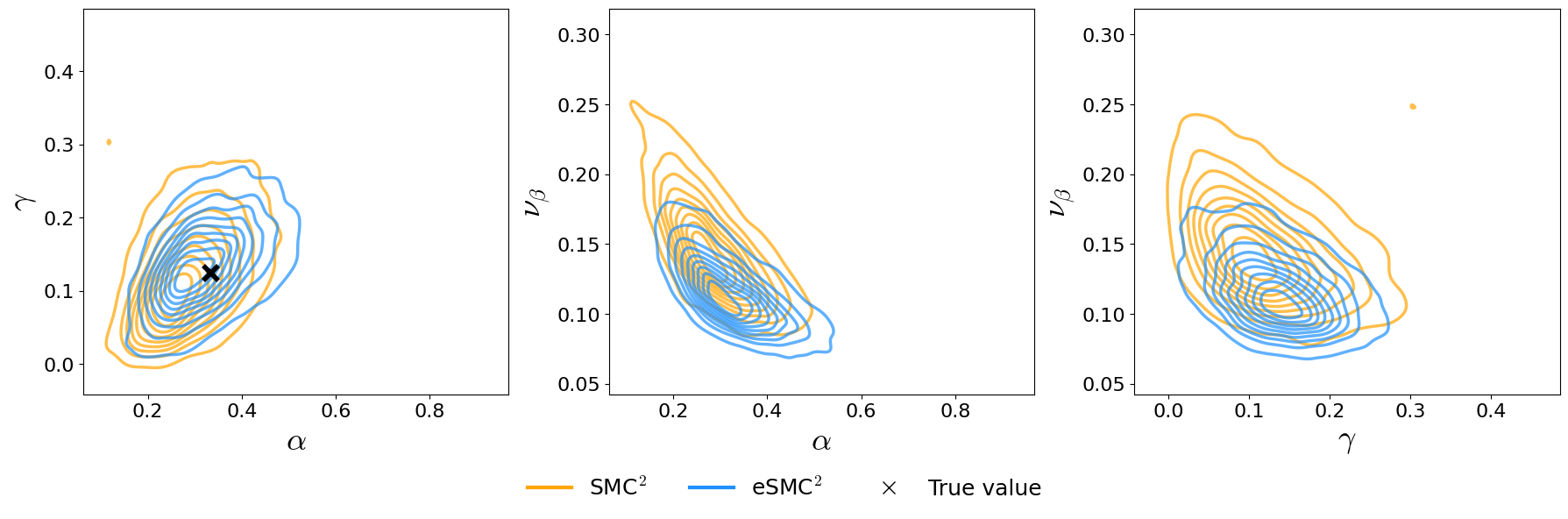}
\caption{\footnotesize \textbf{Posterior distributions of model parameters for Example 2.} Each subplot shows the pairwise marginal densities of $(\alpha, \gamma)$, $(\alpha, \nu_{\beta})$, and $(\gamma, \nu_{\beta})$. Contours represent the combined posterior from five runs of SMC$^2$ (orange) and eSMC$^2$ (blue), with black crosses indicating the true parameter values.}

    \label{fig:contour2}
\end{figure}

\begin{figure}[H]
    \centering
    \includegraphics[width=0.8\linewidth]{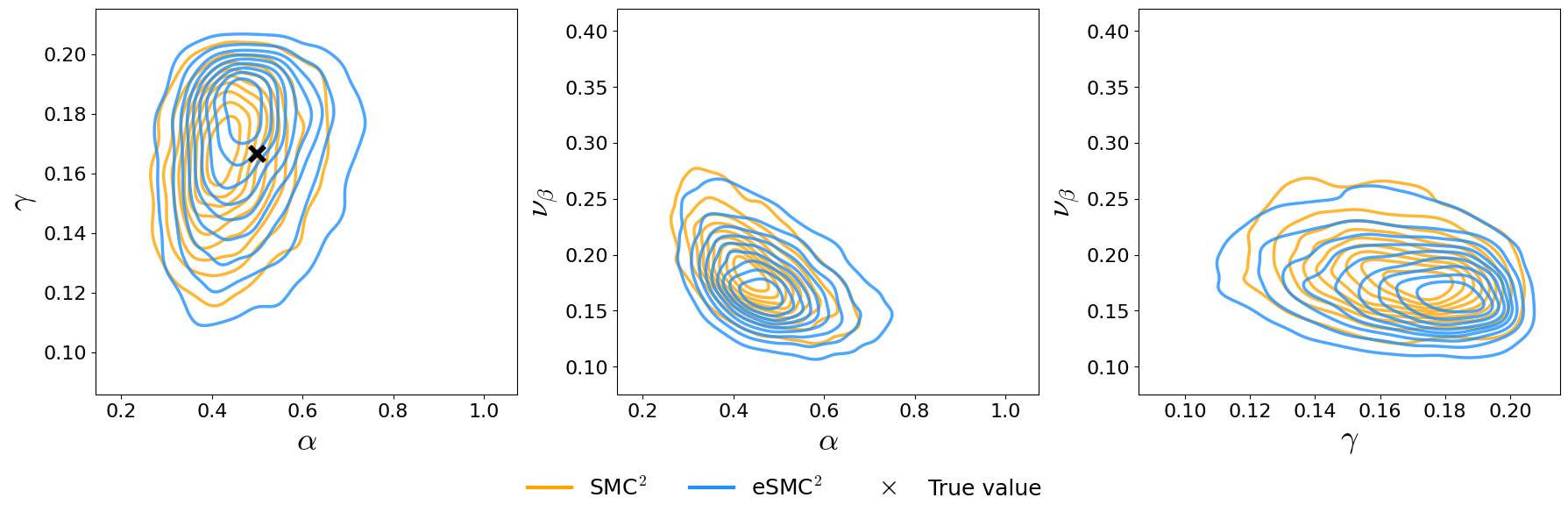}
\caption{\footnotesize \textbf{Posterior distributions of model parameters for Example 3.} Each subplot shows the pairwise marginal densities of $(\alpha, \gamma)$, $(\alpha, \nu_{\beta})$, and $(\gamma, \nu_{\beta})$. Contours represent the combined posterior from five runs of SMC$^2$ (orange) and eSMC$^2$ (blue), with black crosses indicating the true parameter values.}

    \label{fig:contour3}
\end{figure}

This section presents supplementary figures supporting the results discussed in the main text. 
Figures~\ref{fig:unobs1}--\ref{fig:unobs3} show the filtered estimated trajectories of the unobserved SEIR compartments, 
while Figures~\ref{fig:contour1}--\ref{fig:contour3} display the posterior pairwise parameter distributions obtained 
from SMC$^2$ and eSMC$^2$ for Examples~1,~2 and~3.

\section{Sensitivity to ensemble size and prior}\label{appD}
\begin{figure}[H]
    \centering
    \includegraphics[width=0.8\linewidth]{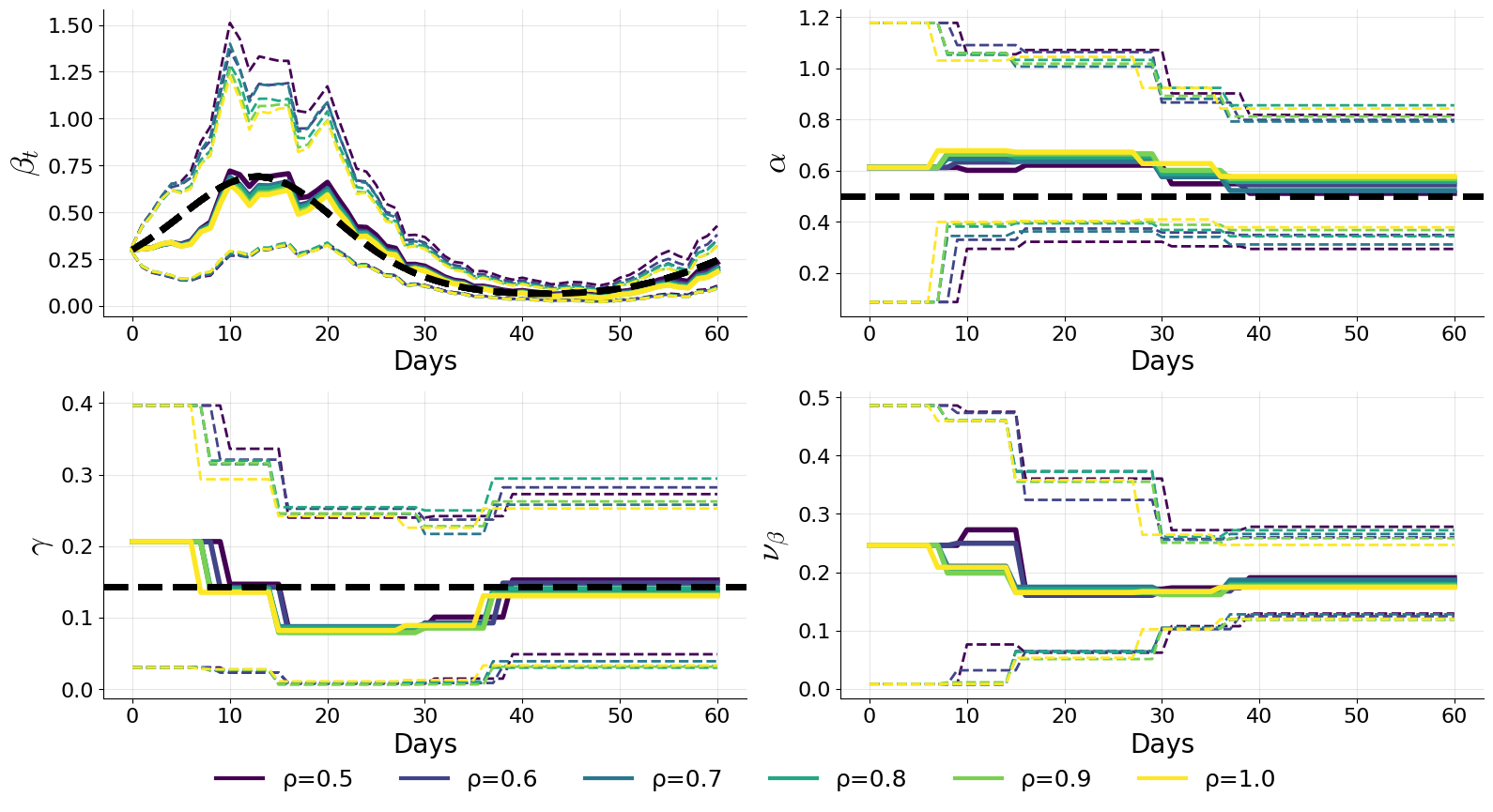}
 \caption{\footnotesize \textbf{Filtered estimate estimates of  $\beta_t$, $\alpha$, $\gamma$, and $\nu_{\beta}$}. Results are obtained using the eSMC$^2$ algorithm on Example 1 under different fixed reporting fractions $\rho$. Solid lines denote posterior medians and dashed lines the associated 95\% credible intervals; black dashed lines indicate the ground truth value.}
    \label{fig:rho_sens}
\end{figure}

Table~\ref{tab:esmc2_nx} reports the results of eSMC$^2$ for different values of the number of state particles ($N_x$), with the number of parameter particles fixed at $N_{\theta} = 1000$.
 Across both examples, increasing $N_x$ leads to higher computational cost but only marginal improvements in parameter estimate, suggesting that moderate ensemble sizes (around $N_x = 150-300$) offer a good trade-off between accuracy and efficiency. 
\begin{table}[H]
\small
\centering
\caption{\footnotesize \textbf{eSMC$^2$ sensitivity analysis across $N_x$.} Posterior mean and standard deviation (in parentheses) are shown for each estimated parameter. CPU time is reported in seconds.}
\label{tab:esmc2_nx}
\begin{tabular}{ccccc}
\toprule
$\bm{N_x}$ & \textbf{CPU} & $\bm{\alpha}$ & $\bm{\gamma}$ & $\bm{\nu_{\beta}}$ \\
\midrule

\multicolumn{5}{l}{\textbf{\textit{Example 1}}} \\
50  & 271  & 0.589 (0.089) & 0.161 (0.056) & 0.174 (0.026) \\
100 & 302  & 0.580 (0.126) & 0.161 (0.048) & 0.169 (0.030) \\
200 & 430  & 0.560 (0.126) & 0.142 (0.055) & 0.178 (0.038) \\
400 & 986  & 0.554 (0.107) & 0.162 (0.055) & 0.171 (0.027) \\
800 & 1320 & 0.558 (0.111) & 0.154 (0.055) & 0.178 (0.030) \\
\multicolumn{2}{l}{\textit{Truth}} & 0.5 & 0.142 & - \\

\midrule

\multicolumn{5}{l}{\textbf{\textit{Example 2}}} \\
50  & 360  & 0.294 (0.055) & 0.123 (0.045) & 0.129 (0.019) \\
100 & 401  & 0.311 (0.085) & 0.132 (0.050) & 0.123 (0.023) \\
200 & 627  & 0.304 (0.068) & 0.131 (0.041) & 0.120 (0.020) \\
400 & 1035 & 0.302 (0.076) & 0.143 (0.052) & 0.129 (0.027) \\
800 & 1770 & 0.314 (0.080) & 0.116 (0.056) & 0.131 (0.027) \\
\multicolumn{2}{l}{\textit{Truth}} & 0.333 & 0.125 & - \\

\midrule

\multicolumn{5}{l}{\textbf{\textit{Example 3}}} \\
50 & 304 & 0.479 (0.075) & 0.167 (0.017) & 0.169 (0.022) \\
100 & 379 & 0.484 (0.086) & 0.169 (0.019) & 0.166 (0.025) \\
200 & 615 & 0.479 (0.078) & 0.173 (0.016) & 0.162 (0.021) \\
400 & 1000 & 0.483 (0.102) & 0.166 (0.021) & 0.176 (0.033) \\
800 & 1921 & 0.485 (0.094) & 0.166 (0.021) & 0.179 (0.032) \\
\multicolumn{2}{l}{\textit{Truth}} & 0.5 & 0.167 & - \\

\bottomrule
\end{tabular}
\end{table}
In Figure~\ref{fig:rho_sens}, we examine how changes in the reporting fraction $\rho$ affect the recovery of key model parameters. The simulated data were generated under the same settings as in Example~1, but with reporting probabilities $y_t \sim \text{Poiss}(\rho Z_t)$ for $\rho \in \{0.5,0.6,0.7,0.8,0.9,1\}$. During inference, we assumed that $\rho$ was known and matched the value used in data generation. The results show that, although the magnitudes of some estimates, especially $\beta_t$ and the early values of $\gamma$ and $\nu_{\beta}$, vary with $\rho$, the overall temporal patterns and qualitative behaviour remain largely robust across reporting scenarios. This indicates that the framework can reliably recover the main dynamic features of the system when the reporting fraction is correctly specified.

\begin{figure}[H]
    \centering
    \includegraphics[width=0.8\linewidth]{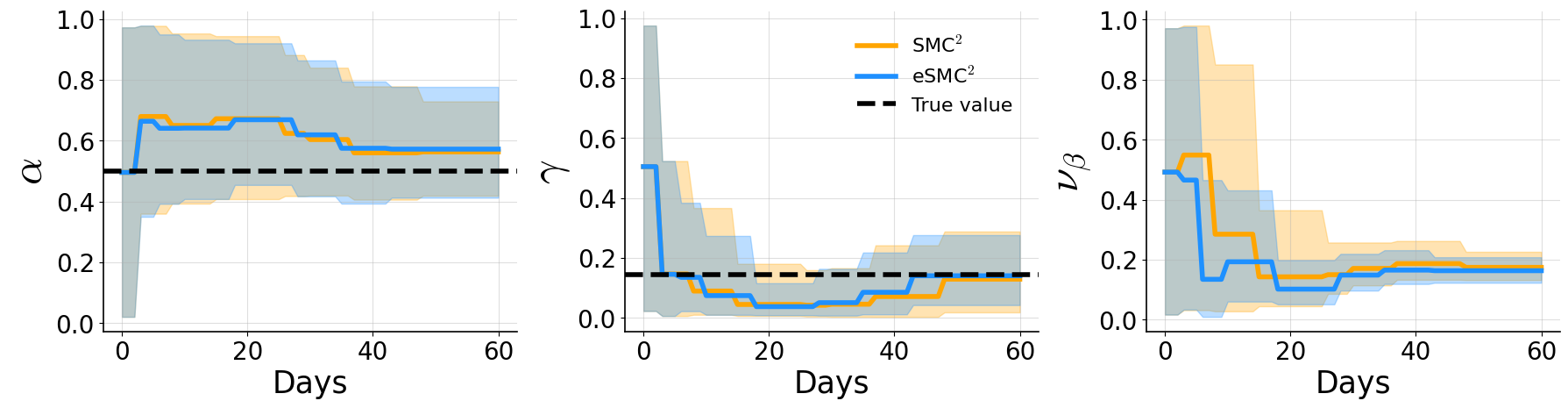}
    \caption{\footnotesize \textbf{Example 1: Sensitivity of posterior estimates to prior specification.} filtered means with 95\% credible intervals (top) and pairwise marginal
distributions at the final time step ($T=60$) (bottom). The black dashed lines denote the true parameter values.}
    \label{fig:prior_sens1}
\end{figure}

\begin{figure}[H]
    \centering
    \includegraphics[width=0.8\linewidth]{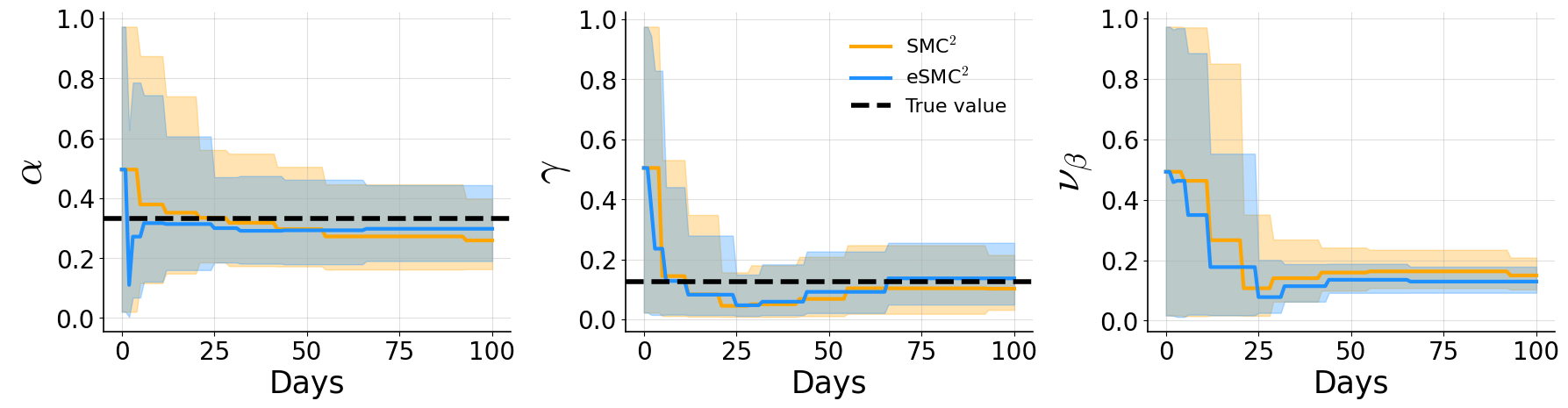}
    \caption{\footnotesize \textbf{Example 2: Sensitivity of posterior estimates to prior specification.} filtered means with 95\% credible intervals (top) and pairwise marginal
distributions at the final time step ($T=100$) (bottom). The black dashed lines denote the true parameter values.}
    \label{fig:prior_sens2}
\end{figure}

\begin{figure}[H]
    \centering
    \includegraphics[width=0.8\linewidth]{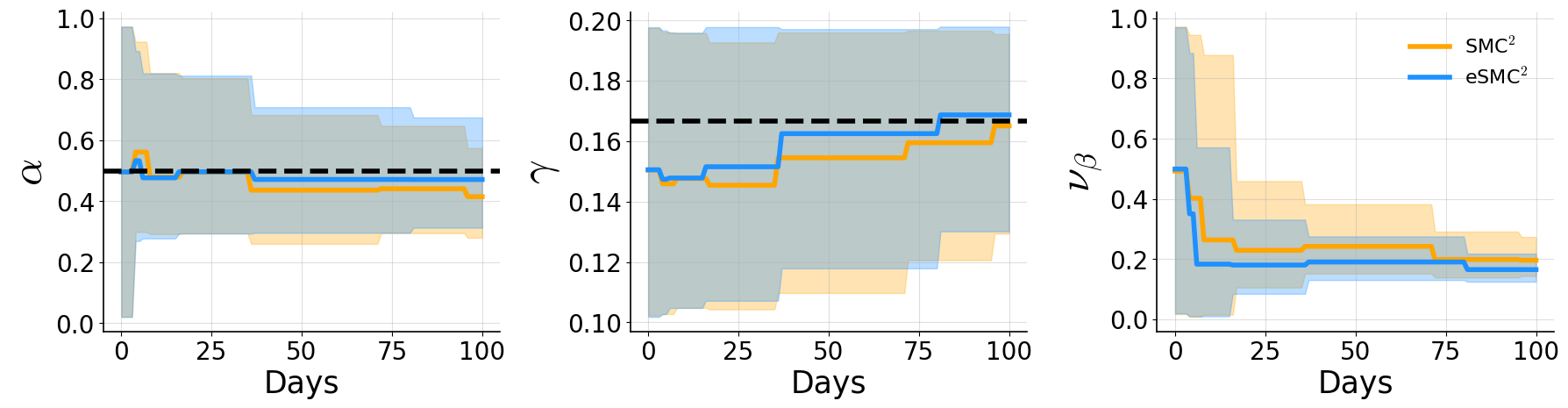}
    \caption{\footnotesize \textbf{Example 3: Sensitivity of posterior estimates to prior specification.} filtered means with 95\% credible intervals (top) and pairwise marginal
distributions at the final time step ($T=100$) (bottom). The black dashed lines denote the true parameter values.}
    \label{fig:prior_sens3}
\end{figure}

We also assessed the impact of using non-informative priors by assigning $\alpha, \nu_{\beta} \sim \mathcal{U}(0,1)$ and $\gamma \sim \mathcal{U}(0,1)$ for Examples~1 and~2, $\gamma \sim \mathcal{U}(0.1, 0.2)$ for Example~3,  and fitting the model to simulated data. With flat priors, the posterior distribution is largely determined by the likelihood, causing the posterior mode to align closely with the maximum likelihood estimate. Figures~\ref{fig:prior_sens1},~\ref{fig:prior_sens2} and~\ref{fig:prior_sens3} summarize the results for Examples~1,~2 and~3, respectively, showing the evolution of the filtered parameter trajectories and the final marginal and joint posterior distributions.

In both examples, the filtered means for $\alpha$ and $\gamma$ quickly converge toward the true parameter values, and the $95\%$ credible intervals narrow as more data are assimilated, indicating progressive information gain from the likelihood. At the final time step, the marginal posterior densities and contour plots reveal compact, unimodal surfaces centered near the true values. The parameters $\alpha$ and $\gamma$ are particularly well recovered. However, the posterior of $\nu_{\beta}$ exhibits a slight shift relative to SMC$^2$. Crucially, the corresponding contour plots show no qualitative change in the joint-dependence structure, that is, eSMC$^2$ does not produce different correlations or spurious multimodality. 

\section{Comparison with the Liu and West filter}\label{appE}

To further evaluate the performance of eSMC$^2$, we compare it with the well-established particle filtering approach of \citet{liu2001combined}, commonly known as the Liu and West filter. This algorithm provides a benchmark for joint sequential state and parameter estimation within state-space models. It extends the standard particle filter by augmenting the state vector with static parameters and applying kernel density shrinkage to mitigate parameter degeneracy. The algorithm generates samples $\{x_t^{(i)}, \theta_t^{(i)}\}_{i=1}^{N_x} \sim p(x_t, \theta | y_{1:t})$.

At time $t-1$, let $\{x_{t-1}^{(i)}, \theta_{t-1}^{(i)}\}_{i=1}^{N_x}$ denote a set of particles with associated normalized weights $\{W_{t-1}^{(i)}\}_{i=1}^{N_x}$ that collectively approximate the joint posterior distribution $p(x_{t-1}, \theta | y_{1:t-1})$. Upon receipt of a new observation $y_t$, the aim is to update this representation to approximate the posterior $p(x_t, \theta| y_{1:t})$. This distribution can be written recursively as
\begin{align}
p(x_t, \theta | y_{1:t}) &\propto p(y_t | x_t, \theta)\, p(x_t, \theta | y_{1:t-1}) \notag\\
&\propto p(y_t | x_t, \theta)\, p(x_t | y_{1:t-1}, \theta)\, p(\theta | y_{1:t-1}).   
\end{align}
The Liu and West filter extends the standard particle filter by incorporating parameter learning through kernel density shrinkage, thereby mitigating the degeneracy problem that arises when static parameters are treated as fixed latent states. At each time step, the parameter posterior distribution is approximated by a Gaussian mixture:
\begin{align}\label{eq:liu_west_posterior}
p(\theta | y_{1:t-1}) \approx \sum_{i=1}^{N_x} 
W_{t-1}^{(i)}\, \mathcal{N}\!\big(\theta ;\, \lambda\, \theta_{t-1}^{(i)} + (1 - \lambda)\, \bar{\theta}_{t-1},\, h^2 V_{t-1}\big),
\end{align}
where  $\bar{\theta}_{t-1}$ and $V_{t-1}$ denote the empirical mean and covariance of the posterior samples 
$\{\theta_{t-1}^{(i)}, W_{t-1}^{(i)}\}_{i=1}^{N_x}$ at time $t-1$.

The shrinkage parameter $\lambda$ is defined as $\lambda = \sqrt{1 - h^2}$, where  
$h^2 = 1 - \left(\frac{3\delta - 1}{2\delta}\right)^2$ is the kernel smoothing parameter.  
This shrinkage step centers each kernel component around the global mean $\bar{\theta}_{t-1}$ and ensures that the mixture variance matches the posterior variance, thereby preventing over-dispersion of the particle cloud.  
Following \citet{liu2001combined}, we set $\delta = 0.99$ and used $N_x = 20{,}000$ particles.  
The prior specifications are identical to those described in Section~3 of the main text.

\begin{figure}[H]
    \centering
    \includegraphics[width=1\linewidth]{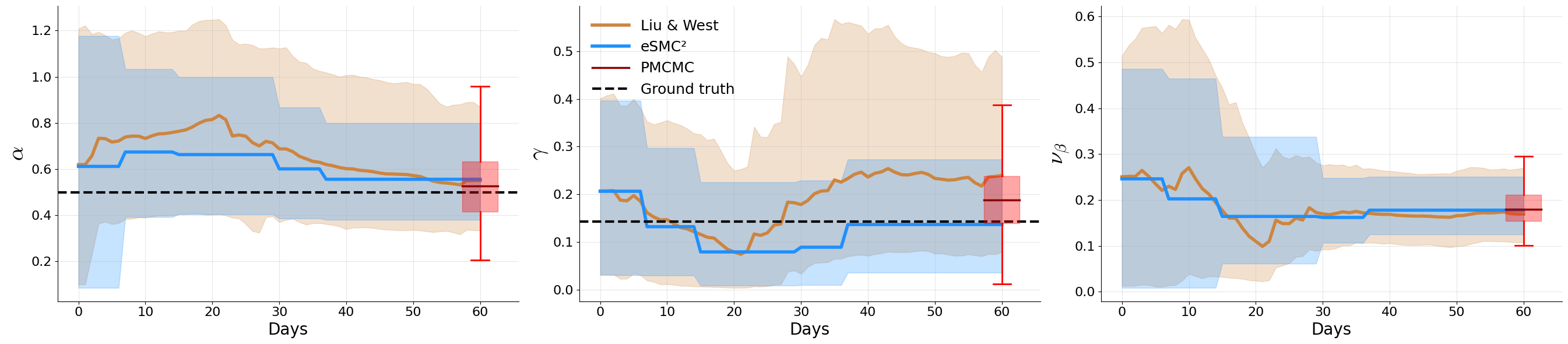}
    \caption{\footnotesize 
    \textbf{Comparison between the Liu and West filter, eSMC$^2$, and PMCMC for simulated data in Example 1.} 
    Filtered means and $95\%$ credible intervals over time for parameters $\alpha$, $\gamma$, and $\nu_{\beta}$. Boxplots at $T=60$ display the marginal posterior distributions obtained from the PMCMC at time 60.}
    \label{fig:liu_west1}
\end{figure}

Figure~\ref{fig:liu_west1} compares filtered estimates obtained with the Liu and West filter and eSMC$^2$, along with the final posterior distributions from a long PMCMC run 
($10^4$ iterations), displayed as boxplots at $T=60$. Both filtering methods track the parameter trajectories, but differences arise in accuracy and uncertainty quantification. 
The Liu and West filter consistently overestimates the recovery rate $\gamma$, likely due to kernel-based shrinkage introducing bias in the joint estimation of static parameters. 
It also produces wider and less stable credible bands, reflecting increased Monte Carlo variability and weight degeneracy. In contrast, eSMC$^2$ yields posterior means closer to the true values with narrower, more coherent uncertainty intervals, highlighting its improved stability and accuracy in sequential updating. The PMCMC boxplots serve as a 
gold-standard reference: eSMC$^2$ credible intervals align closely with the PMCMC posterior at $T=60$, whereas the Liu and West filter shows more pronounced discrepancies, 
particularly for $\gamma$.

\section{Additional results on the mpox dynamics}\label{appF}

This section presents supplementary results supporting the analyses in the main text. We first illustrate inference under a Poisson observation model, then present additional results under a Negative Binomial observation model, and finally discuss forecasting performance comparing eSMC$^2$ with a simpler autoregressive baseline.

\subsection*{Inference under Poisson observation variance}

\begin{figure}[H]
    \centering
    \includegraphics[width=0.85\linewidth]{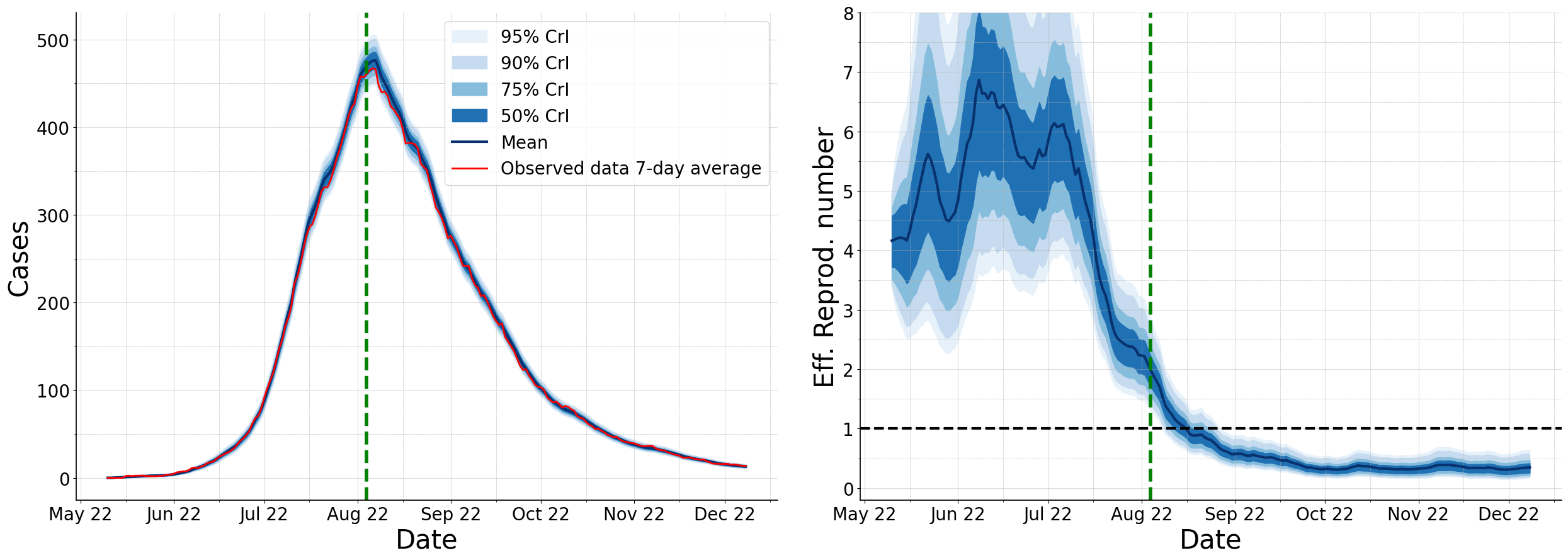}
\caption{\footnotesize \textbf{Inference of weekly incidence and effective reproduction number under Poisson observation variance.} Filtering mean and 95\% credible intervals are shown for reported mpox incidence and estimated $R_t$.}
\label{fig:poiss_mpox}
\end{figure}

Figure~\ref{fig:poiss_mpox} shows estimated weekly incidence and the effective reproduction number for the 2022 U.S. mpox outbreak assuming a Poisson observation model.  
The filtering procedure captures the overall epidemic trajectory under this simplified variance assumption.

\subsection*{Additional results under Negative Binomial observation variance}

\begin{figure}[H]
    \centering
    \includegraphics[width=0.85\linewidth]{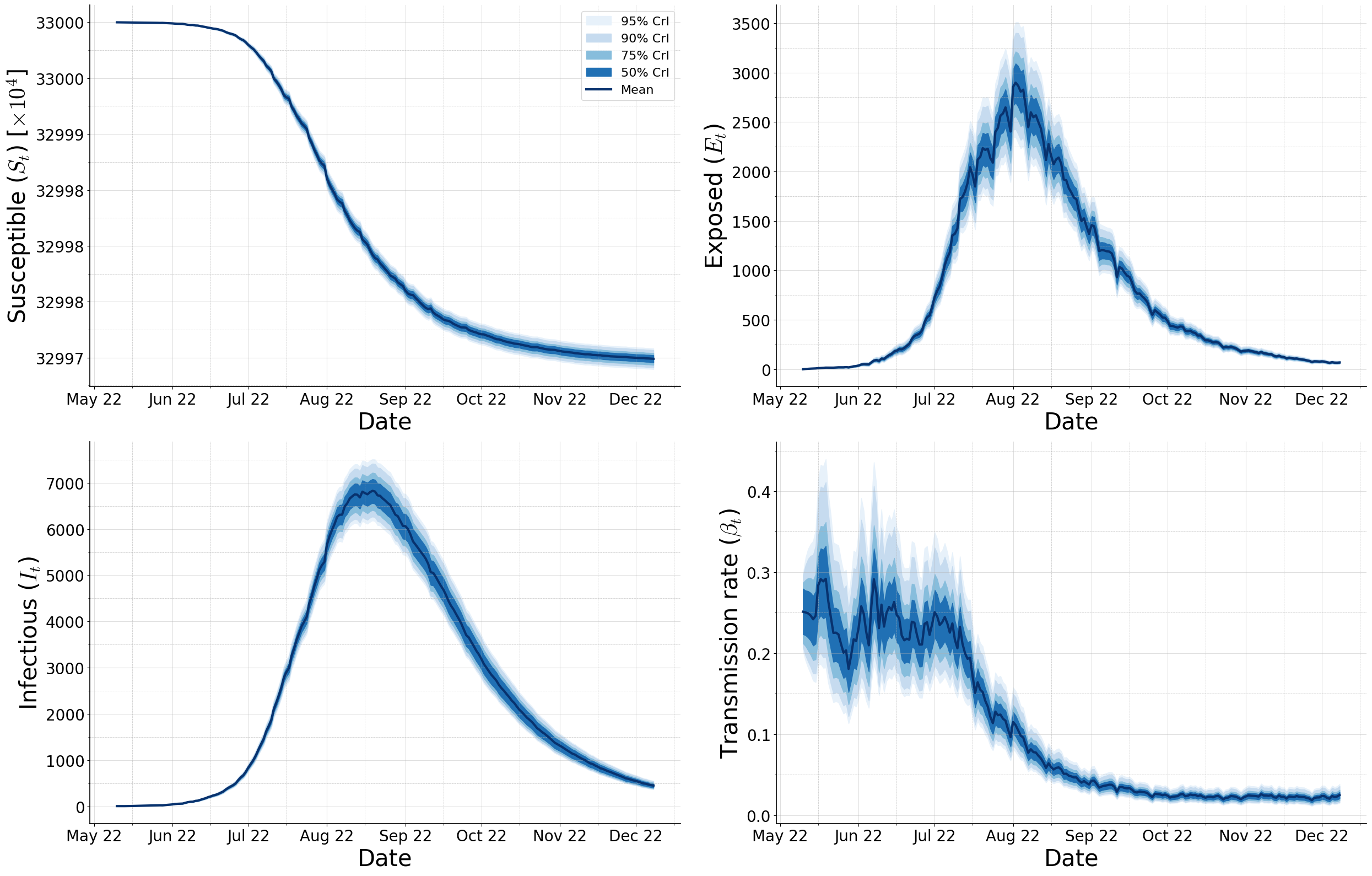}
\caption{\footnotesize \textbf{Unobserved SEIR states under Negative Binomial observation variance.} Filtering mean and 95\% credible intervals are shown for $S_t, E_t, I_t,$ and $\beta_t$.}
\label{fig:latent_mpox}
\end{figure}

\begin{figure}[H]
    \centering
    \includegraphics[width=0.85\linewidth]{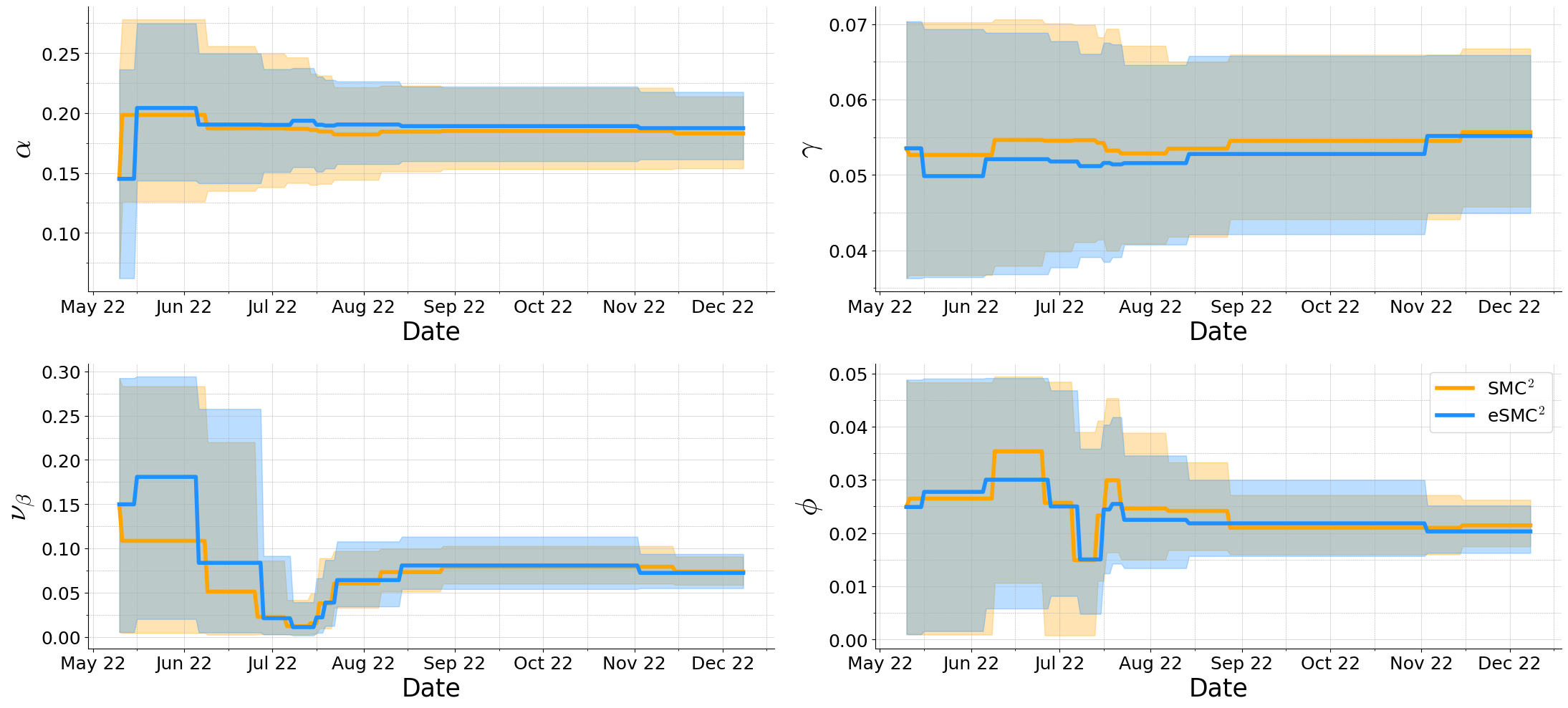}
\caption{\footnotesize \textbf{Comparison of filtered parameter estimates under Negative Binomial observation variance.} Filtered means and 95\% credible intervals obtained with SMC$^2$ and eSMC$^2$ are shown.}
\label{fig:mpox_com_param}
\end{figure}

Figures~\ref{fig:latent_mpox} and \ref{fig:mpox_com_param} present results obtained under a Negative Binomial observation model.  

\subsection*{Autoregressive baseline model for comparison}

For reference, we implemented a discrete-time autoregressive model of order $p$:
\[
y_t = a + \sum_{k=1}^{p} b_k y_{t-k} + \varepsilon_t,
\]
where $Y_t$ is the observed incidence at time $t$, $a$ is an intercept, $b_k$ are autoregressive coefficients, and $\varepsilon_t \sim \mathcal{N}(0, \sigma^2)$ represents the innovation term. Predictive distributions were generated by iteratively simulating forward trajectories conditional on the last $p$ observed values. This baseline captures short-term temporal dependence but does not explicitly model latent transmission dynamics.

\subsection*{Forecast evaluation metrics}

Let $F_{_{T+h}}$ denote the predictive distribution at horizon $_{T+h}$ and $y_{_{T+h}}$ the observed incidence. Forecast performance was evaluated using the following metrics:

\begin{itemize}
    \item \textbf{Mean Absolute Error (MAE)}: 
    \[
        \text{MAE} = \frac{1}{H} \sum_{h=1}^{H} |y_{_{T+h}} - \hat{m}_{_{T+h}}|, 
    \]
    where $H$ is the forecast horizon and $\hat{m}_{_{T+h}}$ is the predictive median.

    \item \textbf{Weighted Interval Score (WIS)}: 
    To jointly assess probabilistic calibration and sharpness, we compute WIS using central $(1-\alpha)$ prediction intervals \citep{bracher2021evaluating}. Let $l_{\alpha, _{T+h}}$ and $u_{\alpha, _{T+h}}$ be the lower and upper bounds of the $(1-\alpha)$ interval from $F_{_{T+h}}$. The interval score at horizon $_{T+h}$ is
    \[
    \resizebox{\textwidth}{!}{$
    \text{IS}_{\alpha}(F_{_{T+h}},y_{_{T+h}}) = (u_{\alpha, _{T+h}} - l_{\alpha, _{T+h}}) 
    + \frac{2}{\alpha} (l_{\alpha, _{T+h}} - y_{_{T+h}}) \mathbf{1}\{y_{_{T+h}} < l_{\alpha, _{T+h}}\} 
    + \frac{2}{\alpha} (y_{_{T+h}} - u_{\alpha, _{T+h}}) \mathbf{1}\{y_{_{T+h}} > u_{\alpha, _{T+h}}\}
    $}
    \]
    The WIS combines $K$ intervals and the absolute error of the median:
    \[
    \text{WIS} =  \frac{1}{H}\sum_{h=1}^{H} 
    \left[\frac{1}{K + 1/2}\left(w_0 |y_{_{T+h}}-\hat{m}_{_{T+h}}| + \sum_{k=1}^{K} w_k \text{IS}_{\alpha_k}(F_{_{T+h}},y_{_{T+h}})\right)\right],
    \]
    where $w_0 = 1/2$ and $w_k = \alpha_k / 2$ are the weights of individual interval scores.

\item \textbf{Empirical 95\% prediction interval coverage}:  To measure the proportion of observed outcomes that fall within the model's 95\% predictive interval, defined as
    \[
    \text{Coverage}_{95} = \frac{1}{H} \sum_{h=1}^{H} \mathbf{1}\{y_{_{T+h}} \in [l_{0.95, _{T+h}}, u_{0.95, _{T+h}}]\},
    \]
    where $l_{0.95, _{T+h}}$ and $u_{0.95, _{T+h}}$ are the lower and upper bounds of the 95\% prediction interval at horizon $_{T+h}$. 
\end{itemize}
\subsection*{Long-term forecast performance using eSMC$^2$}

\begin{figure}[H]
    \centering
    \includegraphics[width=0.8\linewidth]{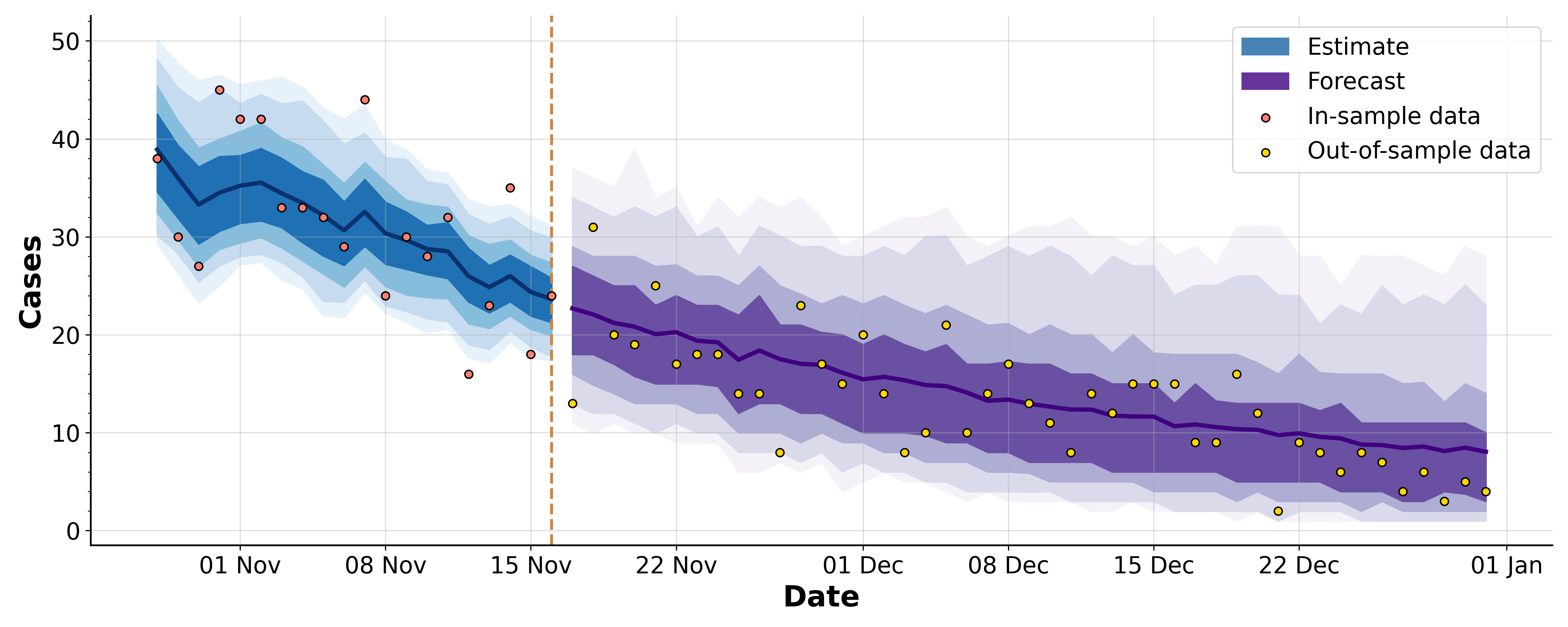}
\caption{\footnotesize \textbf{Long-term forecast performance with eSMC$^2$.} Forecast distributions with 50\%, 75\%, 90\%, and 95\% credible intervals are shown in purple. In-sample observations (red) occur before the forecast start date (vertical dashed line), while out-of-sample observations (yellow) are used for validation.}
\label{Fig10}
\end{figure}

Figure~\ref{Fig10} presents forecasts for the final outbreak phase. The model captures the overall declining trend, though short-term daily fluctuations are less well predicted. Predictions are well calibrated, with 95\% posterior credible intervals covering all out-of-sample observations. These results confirm that eSMC$^2$ produces reliable probabilistic forecasts for real-time epidemic monitoring.

\bibliographystyle{agsm}
\bibliography{references}

\end{document}